\documentclass[10pt,journal,twoside]{IEEEtran}
\usepackage{multirow}
\usepackage{amssymb}
\usepackage[dvips]{graphicx}
\usepackage{amsmath}
\usepackage{amsthm}
\usepackage{latexsym,bm}
\usepackage{color}
\usepackage{subfigure}
\usepackage{longtable}
\usepackage{accents}
\usepackage{cite}
\usepackage{enumerate}
\usepackage{arydshln}
\usepackage[]{pifont}
\usepackage{cuted}
\usepackage{bm}
\usepackage{setspace}
\usepackage{booktabs}
\usepackage{diagbox}
\usepackage{array, caption, threeparttable}
\usepackage{ragged2e}
\usepackage{booktabs,makecell, multirow, tabularx}
\usepackage[ruled]{algorithm}
\usepackage{algorithmic}
\usepackage{tabu}                     
\usepackage{multirow}                 
\usepackage{multicol}                 
\usepackage{multirow}                
\usepackage{float}                    
\usepackage{makecell}                 
\usepackage{booktabs}                 

\usepackage{algorithmic}
\usepackage{algorithm}

\usepackage{diagbox}

\makeatletter
\def\widebar{\accentset{{\cc@style\underline{\mskip10mu}}}}
\def\Widebar{\accentset{{\cc@style\underline{\mskip8mu}}}}
\makeatother

\allowdisplaybreaks
\theoremstyle{plain}

\theoremstyle{definition}
\theoremstyle{definition} 

\begin{document}

\title{STAR-RIS Aided MISO SWIPT-NOMA System with Energy Buffer: Performance Analysis and Optimization
\thanks{K. Xie and G. Cai are with the School of Information Engineering, Guangdong University of Technology, China (e-mail: xiekengyuan@126.com, caiguofa2006@gdut.edu.cn).}
\thanks{J. He is with the Technology Innovation Institute, Abu Dhabi, United Arab Emirates, and also with Centre for Wireless Communications, University of Oulu, Oulu 90014, Finland (e-mail: jiguang.he@tii.ae).}
\thanks{G. Kaddoum is with the Department of Electrical Engineering, University of Qu$\acute{\mathrm{e}}$bec, Qu$\acute{\mathrm{e}}$bec, QC G1K 9H7, Canada, and also with the LaCIME Laboratory, $\acute{\mathrm{E}}$cole de Technologie Sup$\acute{\mathrm{e}}$rieure ($\acute{\mathrm{E}}$TS), Montreal, QC H3C 1K3, Canada (e-mail: georges.kaddoum@etsmtl.ca).}
}
\author{Kengyuan Xie, Guofa Cai, Jiguang He, {\em Senior Member, IEEE}, Georges Kaddoum, \IEEEmembership{Senior Member, IEEE} }

\maketitle
\begin{abstract}
In this paper, we propose a simultaneous transmitting and reflecting reconfigurable intelligent surface (STAR-RIS) and energy buffer aided multiple-input single-output (MISO) simultaneous wireless information and power transfer (SWIPT) non-orthogonal multiple access (NOMA) system, which consists of a STAR-RIS, an access point (AP), and reflection users and transmission users with energy buffers.
In the proposed system, the multi-antenna AP can transmit information and energy to several single-antenna reflection and transmission users simultaneously by the NOMA fashion in the downlink, where the power transfer and information transmission states of the users are modeled using Markov chains.
The reflection and transmission users harvest and store the energy in energy buffers as additional power supplies, which are partially utilized for uplink information transmission.
The power outage probability, information outage probability, sum throughput, and joint outage probability closed-form expressions of the proposed system are derived over Nakagami-$m$ fading channels, which are validated via simulations.
Results demonstrate that the proposed system achieves better performance as compared to the proposed system with discrete phase shifts, the STAR-RIS aided MISO SWIPT-NOMA buffer-less, conventional RIS and energy buffer aided MISO SWIPT-NOMA, and STAR-RIS and energy buffer aided MISO SWIPT-time-division multiple access (TDMA) systems.
Furthermore, a particle swarm optimization-based power allocation (PSO-PA) algorithm is designed to maximize the uplink sum throughput with a constraint on the uplink joint outage probability and Jain's fairness index (JFI).
Simulation results illustrate that the proposed PSO-PA algorithm can achieve an improved sum throughput performance of the proposed system.

\end{abstract}

\vspace{-0.55cm}
\section{Introduction} \label{sect:review}
Internet of Things (IoT) has been considered one of the key and enabling paradigms for a plethora of applications in healthcare, smart cities, and smart homes\cite{9905718}. In the IoT era, various low-power wireless sensor devices are connected to the network and interact with each other.
However, the ever-increasing of wireless electronics inevitably imposes challenges on the power supply of devices and massive multiple access, owing to the limited battery lifespans of devices and scarce spectral resources\cite{10002868}.

Regarding the limited battery challenge although several energy harvesting methods, such as
solar and wind energy harvesting, can capture energy from the environment, their availability is out of control\cite{8758981}. To prolong the operational lifetime of the energy-constrained and low-power devices, simultaneous wireless information and power transfer (SWIPT) has been proposed, which offers a controllable way to charge devices by utilizing the ubiquitous radio-frequency (RF) signals\cite{10032267}. SWIPT enables devices to receive power and information simultaneously.
In addition, to tackle the aforementioned massive connectivity issue, non-orthogonal multiple access (NOMA) technology has been proposed\cite{7263349}. In contrast to orthogonal multiple access (OMA), NOMA allows multiple users to share the same time-frequency resource, therefore providing better spectrum efficiency\cite{7676258}.
Recent studies have combined NOMA and SWIPT to prolong the lifetime of energy-constrained IoT networks and support massive connectivity\cite{9062605,9091839,8485639}.
In \cite{9062605}, a downlink coordinated multipoint SWIPT-NOMA network was proposed.
To further improve the performance, multiple antennas have been introduced into the SWIPT-NOMA system.
In \cite{9091839}, a multiuser multiple-input single-output (MISO) SWIPT-NOMA network was studied, where a novel hybrid user pairing scheme was proposed.
In \cite{8485639}, a millimeter-wave multiple-input multiple-output (MIMO) SWIPT-NOMA system was investigated.
The above works assumed that all harvested energy is consumed in one time slot and energy harvesting is performed with the harvest-use (HU) architecture.
However, in practice, the harvested energy is small and random, which limits the performance of SWIPT-NOMA systems. Therefore, the energy buffer with the harvest-store-use (HSU) architecture was considered for storing the harvested energy, which can improve the performance of self-sustaining nodes\cite{8932568,9712644,9497723}.
In \cite{8932568}, a wireless-powered cooperative NOMA relay network was considered, where the relay was provisioned with an energy buffer to improve energy efficiency.
In \cite{9712644}, an adaptive multiuser cooperative NOMA scheme was studied, where the near-user was equipped with an energy buffer to harvest energy from the ambiance, thus improving the system throughput.
In \cite{9497723}, a MISO wireless-powered NOMA network was investigated, which considers the utilization of a data buffer and energy storage at sources to achieve both a higher sum rate and improved fairness.

Recently, reconfigurable intelligent surfaces (RISs) have received extensive attention from both academia and industry, which can bring enhanced spectrum and energy efficiencies\cite{9398559}.
A RIS is composed of massive reflecting elements that are energy efficient and
cost-effective\cite{9326394}, which has been introduced into SWIPT-NOMA systems for improved system performance.
On the one hand, the efficiency of SWIPT highly depends on the existence of a line-of-sight (LoS) link\cite{7106496} which may be blocked by various obstacles due to long distance transmissions. On the other hand, the far-field user in NOMA may experience poor performance due to path loss.
RISs are not only capable of building virtual LoS links when LoS links are blocked\cite{10032536}, but can also bring outstanding performance gains in SWIPT\cite{8941080}.
Hence, tremendous research efforts have been devoted to RIS aided SWIPT-NOMA systems\cite{9990572,9771850,10254294, 10511286, 10286271, 10319410, 9926196}.
A RIS assisted cooperative SWIPT-NOMA system was investigated in \cite{9990572,9771850}, where \cite{9990572} focused on improving the achievable rate of the strong user while guaranteeing the weak user's quality of service, and \cite{9771850} is devoted to maximizing the data rate of the cell-edge user.
The age of information minimization problem of the RIS aided SWIPT-NOMA system was investigated in \cite{10254294}.
The max-min energy efficiency problem of the active RIS aided multi-cluster cooperative SWIPT-NOMA system was studied in \cite{10511286}.
A RIS aided SWIPT-wireless-powered communication network NOMA system was found to ameliorate transmission
efficiency in \cite{10286271}.
A cooperative NOMA mobile edge computing (MEC) network aided by RIS and SWIPT was analyzed in \cite{10319410}.
A RIS aided MISO SWIPT-NOMA system was proposed in \cite{9926196}, which is considered a cooperative transmission scheme
to improve the performance of cell-edge users.
However, in \cite{9990572,9771850,10254294, 10511286, 10286271, 10319410, 9926196}, the HU architecture was utilized to harvest energy, where energy harvesting nodes do not buffer the energy for use in future time slots.
Moreover, the conventional RIS only reflects the signal to users, which makes it necessary for the transmitter and receiver to be on the same side.

To overcome the limitation of user distribution in conventional RIS aided systems, the novel concept of simultaneous transmitting and reflecting RIS (STAR-RIS) has been recently proposed \cite{9690478}.
Different from the conventional RIS, STAR-RIS provides full-space service coverage and a more flexible deployment\cite{9834288}, which can not only reflect but also transmit (also known as refract) the incident signal.
Specifically, the incident signal on the STAR-RIS is divided into two parts.
One part is reflected to the same space as the transmitter, i.e., the reflection space, and the other part is transmitted to the opposite space, i.e., the transmission space\cite{9570143}.
Moreover, three practical protocols of STAR-RIS have been proposed in \cite{9570143}, i.e., time-switching, mode-switching, and energy-splitting.
Because the STAR-RIS would divide the incident signal into two portions in the energy-splitting protocol and mode-switching protocol, a multiple access scheme is necessary to distinguish these two parts to decode them successfully
\cite{9808307}.
NOMA has been considered to be a promising multiple-access technique due to its ability to provide enhanced spectral efficiency and connectivity\cite{9740451}.
Recently, STAR-RIS aided NOMA networks have been investigated in several works\cite{9863732,9964251,9956827}.
In \cite{9863732}, a STAR-RIS aided NOMA system was investigated to maximize the achievable sum rate.
In \cite{9964251}, a STAR-RIS assisted downlink MISO-NOMA network was considered for maximizing
the system energy efficiency.
In \cite{9956827}, a STAR-RIS assisted downlink MIMO-NOMA network was considered to investigate the energy-efficient resource allocation.
To study the potential benefits of deploying STAR-RIS in wireless-powered transmission, there have been some initial works on STAR-RIS aided wireless-powered systems\cite{10032506,10086660,666666,10073379}.
In \cite{10032506}, a STAR-RIS aided wireless-powered MEC system was investigated.
In \cite{10086660}, a STAR-RIS assisted wireless-powered IoT network was proposed, which adopts the time division multiple access (TDMA) scheme.
In \cite{666666}, a STAR-RIS aided wireless-powered NOMA system was investigated, where the time-switching protocol and energy-splitting protocol of STAR-RIS are utilized to improve the sum throughput.
In \cite{10073379}, a STAR-RIS aided SWIPT system was proposed, where STAR-RIS employs the time-switching protocol.
However, in \cite{10032506,10086660,666666,10073379}, only single-antenna access points (APs) with the HU architecture were considered.

In light of the aforementioned motivations, we propose a STAR-RIS and energy buffer aided MISO SWIPT-NOMA system in this paper. The HSU architecture for users with energy buffers is adopted to achieve energy management, thus ensuring better performance.
The main contributions of this paper can be summarized as follows:
\begin{itemize}
\item
We consider a STAR-RIS and energy buffer aided MISO SWIPT-NOMA system, where a multi-antenna AP transmits the superimposed signals to several single-antenna users via the direct link and reflection/transmission links assisted by STAR-RIS in the downlink,
and the users transmit their data to the AP in the uplink.
Successive interference cancellation (SIC) is considered for information decoding.
Moreover, the reflection and transmission users are equipped with energy buffers to enhance the system's  sustainability, where the power transfer and information transmission states are modeled using Markov chains.
In this system, the sustainability is studied from the perspectives of both energy acquisition and
consumption.
\item
By exploiting the moment-matching approach, the proposed system's power outage probability, information outage probability, and sum throughput performances are derived in closed-form over Nakagami-$m$ fading channels{\footnote{Nakagami-$m$ fading channels have been adopted in RIS or STAR-RIS aided communication systems due to their  generic property\cite{9915477,9804844}, which can characterize wireless signal propagation in a variety of environments. It should be noted that Nakagami-$m$ fading channels can be reduced into other channels, such as Rayleigh and Rician, by changing the value of $m$\cite{9804844,9406837}.}}.
The power and information outage probabilities are combined into a closed-form joint outage probability.
Simulations are carried out to verify the accuracy of the theoretical analysis and demonstrate that the proposed system offers improved performance compared to baseline schemes.
\item
To achieve the maximization of uplink sum throughput while ensuring a certain uplink joint outage probability and Jain's fairness index (JFI) for users, a low-complexity particle swarm optimization-based power allocation (PSO-PA) algorithm is proposed to find the optimal energy-splitting ratios of the STAR-RIS and the transmit power allocation factors.
Results unveil that the proposed PSO-PA algorithm provides an improved sum throughput performance
for the proposed system.
\end{itemize}

The rest of this paper is organized as follows.
In Section II, we introduce the system model.
In Section III, we analyze the performance of the proposed system.
In Section IV, we propose the resource allocation method.
Simulation results and discussions are summarized in Section V.
Finally, we conclude the paper in Section VI.

\textit{Notations}:
Matrices and vectors are denoted by uppercase bold and lowercase bold, respectively.
${\mathbb{C}^{M \times 1}}$ denotes the space of $M \times 1$ complex vectors.
${j^2} =  - 1$ and $j$ denotes the imaginary unit.
diag($\mathbf{x}$) denotes a diagonal matrix with the elements of vector $\mathbf{x}$ on the main
diagonal.
The complex Gaussian distribution is denoted by $\mathcal{CN}\left( {\mu ,{\sigma ^2}} \right)$ with mean $\mu$ and variance ${{\sigma ^2}}$.
${K_\alpha }\left(  \cdot  \right)$ denotes the modified Bessel function of the second kind.
${I_\alpha }\left( \cdot \right)$  is the $\alpha$-th order Bessel function of the first kind.
$\gamma \left( { \cdot , \cdot } \right)$ is the upper incomplete Gamma function.
$\Gamma \left(  \cdot  \right)$ is the Gamma function.
$\mathrm{E}\left(  \cdot  \right)$ and $\mathrm{Var}\left(  \cdot  \right)$ represent the expectation and variance operations, respectively.
$\max \left( {a,b} \right)$ returns the maximum of $a$ and $b$.

\section{System Model} \label{sect:system model}
This section first describes the proposed system and its downlink signal model, and then presents the energy consumption model and the uplink signal model.

\begin{figure}[t]
\center
\includegraphics[width=3.1in,height=2.2in]{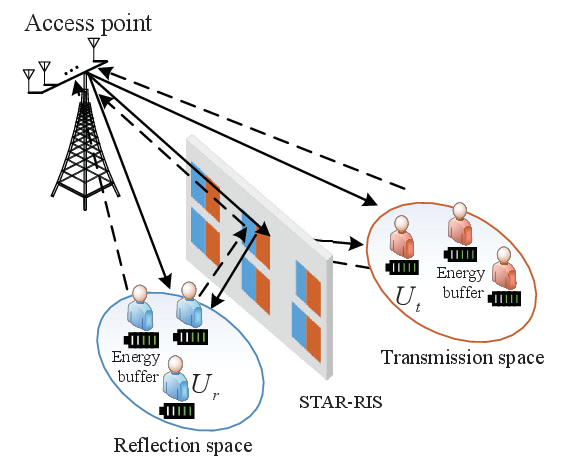}
\vspace{-0.2cm}
\captionsetup{font={footnotesize}}
\caption{STAR-RIS and energy buffer aided MISO SWIPT-NOMA system.}
\label{fig:system}
\vspace{-0.55cm}
\end{figure}
\vspace{-0.5cm}

\subsection{STAR-RIS and Energy Buffer Aided MISO SWIPT-NOMA System}
Fig.~\ref{fig:system} illustrates the proposed STAR-RIS and energy buffer aided MISO SWIPT-NOMA system,
consisting of an $L$-antenna AP, a STAR-RIS, reflection users, and transmission users.
The $L$-antenna AP can communicate with single-antenna reflection users and transmission users with the aid of the STAR-RIS that is composed of $N$  low-cost reflective elements.
It is assumed that the direct link from the AP to the user is available.
In this paper, it is assumed that AP has a fixed energy supply, while ${U_r}$ and ${U_t}$ are powered by harvested energy due to energy constraints.
Time-division duplex (TDD) communication is adopted.
First, the AP simultaneously transmits the energy and information to ${U_r}$ and ${U_t}$ in the downlink using NOMA. Then, ${U_r}$ and ${U_t}$ use the harvested energy to power the signal processing circuit and transmit signals to AP in the uplink.
The information transmission and energy transfer between AP and ${U_k }$, $k  \in \left\{ {t,r} \right\}$ are assisted by STAR-RIS.

{{\em Remark~1:}
Considering the complexity of SIC at the user, NOMA is applied between a pair of users, i.e., two users, which is a basic case studied in previous works\cite{9956827}.
In fact, AP performs user pairing based on the channel gains of different users in NOMA systems\cite{8998131,9091839,8408563}.
In the proposed system, user pairing depends on the channel gains of the users in the reflection and transmission spaces. Specifically, similar to \cite{8998131,9091839,8408563}, the channel gains of the users in the reflection and transmission spaces are ranked from smallest to largest, respectively.
To enlarge the channel gains difference between two users from different spaces, the unmatched user with the highest channel gain in the reflection space is paired with the unmatched user with the lowest channel gain in the transmission space.
Similar to \cite{10223312,9896755}, multiple NOMA-pairs can be served via TDMA to avoid interference between them.
For simplicity of presentation and revealing fundamental
insights, we consider one of NOMA-pairs{\footnote{The derived results can be extended to multiple transmission and reflection users' scenario. In addition, the development of a more efficient user pairing mechanism for multiple users scenarios in STAR-RIS can be an interesting future work.}}, i.e., one reflection user ${U_r}$ and one transmission user ${U_t}$.

}

\vspace{-0.45cm}
\subsection{Downlink Signal Model}
The STAR-RIS works using the energy-splitting protocol{\footnote{In fact, since the mode-switching protocol is a special case of the energy splitting protocol\cite{9570143}, the energy-splitting protocol is considered in this paper.}}, where the AP can simultaneously communicate with users located on opposite sides of the STAR-RIS using NOMA.
In the energy-splitting protocol, each element of the STAR-RIS operates simultaneously in transmission and reflection modes, where $\sqrt {{\beta _r}} ,\sqrt {{\beta _t}}  \in \left[ {0,1} \right]$ denote the energy-splitting ratios.
$\beta_{r} + \beta_{t} = 1$ holds based on the law of energy conservation\cite{9740451}.
Let ${{\boldsymbol{\varphi }}_r} \buildrel \Delta \over = \left[ {\sqrt {{\beta _r}} {e^{j\varphi _1^r}}, \ldots ,\sqrt {{\beta _r}} {e^{j\varphi _N^r}}} \right]$
and ${{\boldsymbol{\varphi }}_t} \buildrel \Delta \over = \left[ {\sqrt {{\beta _t}} {e^{j\varphi _1^t}}, \ldots ,\sqrt {{\beta _t}} {e^{j\varphi _N^t}}} \right]$ denote the STAR-RIS's reflection and transmission coefficient vectors, where $\varphi _i^r,\varphi _i^t \in \left[ {0,2\pi } \right),i \in \left\{ {1,2, \ldots ,N} \right\}$ denote the phase-shift adjustments at the $i$-th element.

\subsubsection{Downlink Information Transmission}
The quasi-static flat-fading channels were assumed\cite{9712644,8811733}.
Let ${\mathbf{\tilde {H}}} \in {\mathbb{C} ^{N \times L}}$ be the \textit{complex channel coefficients} matrix,
${\mathbf{\tilde {g}}_k} \in {\mathbb{C}^{1 \times N}}$ be the \textit{complex channel coefficients} vector, ${{\bf{\tilde{h}}} _k^0} \in {\mathbb{C}^{1 \times L}}$ be the \textit{complex channel coefficients} vector of the direct link between the AP and STAR-RIS, the STAR-RIS and ${U_k}$, and the AP and ${U_k}$.
The AP uses NOMA to form a superimposed signal $s = \sqrt {{P_{\text{AP}}}/L} \left( {\sqrt {{\alpha _r}} {s_r} + \sqrt {{\alpha _t}} {s_t}} \right)$, where ${P_{\text{AP}}}/L$ is the transmit power of each antenna at the AP, ${{\alpha _r}}$ and ${{\alpha _t}}$ are the transmit power allocation factors for ${U_r}$ and ${U_t}$ with ${\alpha _r} + {\alpha _t} = 1$, and ${s_r}$ and ${s_t}$ are the unit-energy information signals of ${U_r}$ and ${U_t}$, i.e., $\mathrm{E}\left( {{{\left| {{s_r}} \right|}^2}} \right) = \mathrm{E}\left( {{{\left| {{s_t}} \right|}^2}} \right) = 1$.

In the proposed system, the power-splitting mode is adopted in the SWIPT, with $\theta$ being the power-splitting factor, which indicates the portion of the received signal used for information decoding.
The received signal for information decoding at ${U_k}$ can be expressed as
\vspace{-0.2cm}
\begin{small}
\begin{align}
\label{eq:yk}
&{y_k} = \sqrt \theta  \left( {{\bf{\tilde h}}_{{k}}^{{0}}{\bf{w}} + {{{\bf{\tilde g}}}_{{k}}}{\rm{diag}}({{\boldsymbol{\varphi }}_{{k}}}){\bf{\tilde Hw}}} \right)s + {n_k}\nonumber\\
& \!=\! \sqrt {\theta {\beta _k}} \!\sum\limits_{\ell  = 1}^L \!{\left( \!{\tilde h_{k,\ell }^0{e^{ - j{\varpi _\ell }}} \!+\! \sum\limits_{i = 1}^N {{{\tilde h}_{\ell ,i}}{{\tilde g}_{k,i}}{e^{j\left( {\varphi _i^k - {\varpi _\ell }} \right)}}} } \!\right)} s \!+\! {n_k},
\vspace{-0.25cm}
\end{align}
\end{small}%
where $\mathbf{w} = \left[ {{e^{ - j{\varpi _1}}}, \ldots ,{e^{ - j{\varpi _L}}}} \right]$ is the beamforming vector, and
${n_{k}}$ is the additive white Gaussian noise (AWGN) with ${n_{k}} \sim {\cal C}{\cal N}\left( {0,{N_0}} \right)$.
\subsubsection{Energy Harvesting}
Both users receive the superimposed signal $s$, which includes the signals ${\sqrt {{\alpha _r}} {s_r}}$ and ${\sqrt {{\alpha _t}} {s_t}}$. Since ${\alpha _r} + {\alpha _t} = 1$, the harvested energy for ${U_k}$ can be calculated as
\vspace{-0.2cm}
\begin{small}
\begin{align}
\label{eq:Er1}
&{E_k} = \eta \left( {1 - \theta } \right){\beta _k}{\left| {\sum\limits_{\ell  = 1}^L {\left( \begin{array}{l}
\tilde h_{k,\ell }^0{e^{ - j{\varpi _\ell }}}\\
 + \sum\limits_{i = 1}^N {{{\tilde h}_{\ell ,i}}{{\tilde g}_{k,i}}{e^{j\left( {\varphi _i^k - {\varpi _\ell }} \right)}}}
\end{array} \right)} } \right|^2}{\alpha _r}\frac{{{P_{\text{AP}}}}}{L}{T_s}\nonumber\\
& + \eta \left( {1 - \theta } \right){\beta _k}{\left| {\sum\limits_{\ell  = 1}^L {\left( \begin{array}{l}
\tilde h_{k,\ell }^0{e^{ - j{\varpi _\ell }}}\\
 + \sum\limits_{i = 1}^N {{{\tilde h}_{\ell ,i}}{{\tilde g}_{k,i}}{e^{j\left( {\varphi _i^k - {\varpi _\ell }} \right)}}}
\end{array} \right)} } \right|^2}{\alpha _t}\frac{{{P_{\text{AP}}}}}{L}{T_s}\nonumber\\
& \!=\! \eta \left( {1 - \theta } \right){\beta _k}{\left| {\sum\limits_{\ell  = 1}^L\! {\left(\! \begin{array}{l}
\tilde h_{k,\ell }^0{e^{ - j{\varpi _\ell }}}\\
 + \sum\limits_{i = 1}^N {{{\tilde h}_{\ell ,i}}{{\tilde g}_{k,i}}{e^{j\left( {\varphi _i^k - {\varpi _\ell }} \!\right)}}}
\end{array} \!\right)} } \right|^2}\frac{{{P_{\text{AP}}}}}{L}{T_s},
\end{align}
\end{small}%
where $\eta$ denotes the energy conversion efficiency factor, and ${T_s}$ is the symbol duration.

The \textit{complex channel coefficients} are represented by
polar coordinates as $\tilde h_{k,\ell }^0 = h_{k,\ell }^0{e^{ - j{\xi _\ell }}}$,
${{\tilde h}_{\ell ,i}} = {h_{\ell ,i}}{e^{ - j{\mu _{\ell ,i}}}}$, ${{\tilde g}_{k,i}} = {g_{k,i}}{e^{ - j{\sigma _i}}}$, where $h_{k,\ell }^0$, ${h_{\ell,i }}$, ${g_{k,i}}$ denote the \textit{ magnitudes
of the channel coefficients} with $\left| {\tilde h_{k,\ell }^0} \right| = h_{k,\ell }^0$, $\left| {{{\tilde h}_{\ell,i}}} \right| = {h_{\ell,i}}$, $\left| {{{\tilde g}_{k,i}}} \right| = {g_{k,i}}$, and ${{\xi _\ell }}$,
${{\mu _{\ell ,i}}}$, ${{\sigma _i}}$, $\left\{ {{{\xi _\ell }},{{\mu _{\ell ,i}}},{\sigma _i}} \right\} \in \left[ {0,2\pi } \right]$, denote the \textit{phases} of $\tilde h_{k,\ell }^0$, ${{\tilde h}_{\ell,i}}$, ${{\tilde g}_{k,i}}$.
The magnitudes of the channel coefficients follow the Nakagami-$m$ distributions. Hence $h_{k,\ell }^0\sim{\rm{Nakagami}}\left( {m_{k,\ell }^0,\Omega _{k,\ell }^0} \right)$, ${h_{\ell,i}}\sim {\rm{Nakagami}}\left( {{m_{\ell,i }},{\Omega _{\ell,i }}} \right)$, ${g_{k,i}}\sim {\rm{Nakagami}}\left( {{m_{k,i}},{\Omega _{k,i}}} \right)$, where $m_{k,\ell }^0 > 0$, ${m_{\ell ,i}} > 0$, ${m_{k,i}} > 0$ denote the shape parameters, and $\Omega _{k,\ell }^0$, ${\Omega _{\ell,i}}$, ${\Omega _{k,i}}$ denote the spread parameters of the Nakagami-$m$ distributions.
The spread parameters are given as $\Omega _{k,\ell }^0 = {l_{{\rm{AP}},{U_k}}}$, ${\Omega _{\ell,i }} = {l_{\mathrm{AP},\mathrm{S\text{-}R}}}$, and ${\Omega _{k,i}} = {l_{\mathrm{S\text{-}R},{U_k}}}$, where ${l_{X,Y}} = \tau /d_{X,Y}^\vartheta $ denotes the path-loss coefficient and is integrated into the spread parameters
with ${{\vartheta}}$ being the path-loss exponent,
${{\tau }}$ the path-loss at a reference distance of $1$ meter, ${d_{X,Y}}$ the distance between $X$ and $Y$ with $X \in \left\{ {{\rm{AP}},S\text{-}R\left( {{\rm{STAR\text{-}RIS}}} \right)} \right\}$ and $Y \in \left\{ {S\text{-}R,{U_t},{U_r}} \right\}$.
Hence, (\ref{eq:Er1}) can be expressed as
\vspace{-0.25cm}
\begin{small}
\begin{align}
\label{eq:Er2}
{E_k} \!=\! \eta \left( {1 - \theta } \right){\beta _k}{\left| {\sum\limits_{\ell  = 1}^L \!{\left(\!\! {\begin{array}{*{20}{l}}
{h_{k,\ell }^0{e^{ - j\left( {{\xi _\ell } + {\varpi _\ell }} \right)}} \!+\! \sum\limits_{i = 1}^N {{h_{\ell ,i}}{g_{k,i}}} }\\
{ \times {e^{j\left( {\varphi _i^k - {\varpi _\ell } - {\mu _{\ell ,i}} - {\sigma _i}} \right)}}}
\end{array}} \!\!\!\right)} } \!\right|^2}\!\!\frac{{{P_{{\rm{AP}}}}}}{L}{T_s}.
\vspace{-0.25cm}
\end{align}
\end{small}%
Considering perfect channel state information (CSI){\footnote{The impact of imperfect CSI on the performance of the proposed system is examined in Sect.~V. In addition, the design of more efficient CSI estimation methods for STAR-RIS is an interesting but challenging future work.}}\cite{9740451,9570143}, the maximum available energy can be obtained by the semidefinite relaxation (SDR) technique with the optimal phase-shift value ${\varpi _\ell }$ of the beamforming vector and $\varphi _i^k$ of the STAR-RIS\cite{8811733}.  To avoid the high computational complexity of SDR, we consider a lower complexity suboptimal scheme. Specifically, first, the phase-shift value ${\varpi _\ell }$ of the beamforming vector is fixed at the AP, where its value is opposite to the phase of the direct link, i.e., ${\varpi _\ell } =  - {\xi _\ell }$.
Then, the maximum available energy can be obtained by the phase-shifts $\varphi _i^k$ of the STAR-RIS, i.e., $\varphi _i^k = {\varpi _\ell } + {\mu _{\ell ,i}} + {\sigma _i}$.
Hence, (\ref{eq:Er2}) can be re-expressed as
\vspace{-0.25cm}
\begin{small}
\begin{equation}
\label{eq:Er3}
{E_k} = \eta \left( {1 - \theta } \right){\beta _k}{\left| {\sum\limits_{\ell  = 1}^L {\left( {h_{k,\ell }^0 + \sum\limits_{i = 1}^N {{h_{\ell ,i}}{g_{k,i}}} } \right)} } \right|^2}\frac{{{P_{\text{AP}}}}}{L}{T_s}.
\vspace{-0.25cm}
\end{equation}
\end{small}%

\vspace{-0.40cm}
\subsection{Energy Consumption Model}
The energy consumption of each user mainly includes two components, i.e., the circuit power and the transmission power consumption\cite{7378860}. Specifically, the circuit power consumption for transmitting and receiving one bit is ${E_c}$~(nJ/bit), while the transmission power consumption per bit is ${E_{tp}}$~( pJ/bit/$\mathrm{{m^2}}$).
Moreover, according to \cite{9464720}, the computing module power consumption
is ${E_{comp}}$~(nJ/bit). Hence, the overall power consumption of the user for receiving and transmitting $M$ bits over a distance ${d_{\text{AP},{U_k}}}$ can be calculated as
\vspace{-0.15cm}
\begin{small}
\begin{equation}
{E_s} = 2M{E_c} + M{d_{\text{AP},{U_k}}^2}{E_{tp}} + {E_{comp}},
\vspace{-0.15cm}
\end{equation}
\end{small}%
where both the circuit power consumption for receiving and transmitting $M$ bits at each user consumes $M{E_c}$ energy. In particular, typical values for ${E_c}$ and ${E_{tp}}$ are $50$~nJ/bit and $100$~pJ/bit/$\mathrm{{m^2}}$, respectively \cite{7378860}. In addition, the power consumption of the computing module is about $8\%$ of the total power consumption\cite{9464720,7446052}, i.e., ${E_{comp}} \approx 8.7\% \left( {2M{E_c} + Md_{\text{AP},{U_k}}^2{E_{tp}}} \right)$.

\vspace{-0.45cm}
\subsection{Uplink Signal Model}
Both users simultaneously send their signals to AP by utilizing the transmission power of the energy consumption model. The received signal at the AP can be expressed as
\vspace{-0.2cm}
\begin{small}
\begin{align}
\label{eq:yAP}
&{y_{AP}} = \sqrt {{P_t}{\beta _t}} \sum\limits_{\ell  = 1}^L {\left( {\tilde h_{t,\ell }^0{e^{ - j{\varpi _\ell }}} + \sum\limits_{i = 1}^N {{{\tilde h}_{\ell ,i}}{{\tilde g}_{t,i}}{e^{j\left( {\varphi _i^t - {\varpi _\ell }} \right)}}} } \right)} {s_t}\nonumber\\
& \!+\! \sqrt {{P_r}{\beta _r}} \sum\limits_{\ell  = 1}^L \!{\left(\! {\tilde h_{r,\ell }^0{e^{ - j{\varpi _\ell }}} + \sum\limits_{i = 1}^N {{{\tilde h}_{\ell ,i}}{{\tilde g}_{r,i}}{e^{j\left( {\varphi _i^r - {\varpi _\ell }} \right)}}} } \!\right)} {s_r} + L{n_k},
\vspace{-0.55cm}
\end{align}
\end{small}%
where ${P_k} = \frac{{Md_{{\rm{AP}},{U_k}}^2{E_{tp}}}}{{{T_s}}}$.

\vspace{-0.35cm}
\section{Performance Analysis} \label{sect:perfomnanceanalysis}
This section presents the proposed system's closed-form power outage probability expression over Nakagami-$m$ fading channels. Moreover, the downlink and uplink closed-form information outage probability and sum throughput expressions are derived.
Based on power and information outage probabilities, the downlink and uplink closed-form joint outage probabilities of the proposed system are obtained.
\begin{figure}
\center
\includegraphics[width=3.4in,height=1.7in]{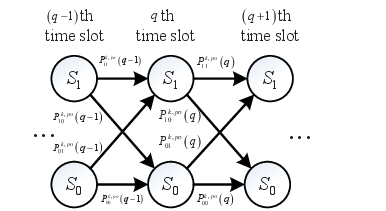}
\vspace{-0.2cm}
\captionsetup{font={footnotesize}}
\caption{The Markov chain model of power states with a rechargeable energy buffer (${S_0}$: Outage, and ${S_1}$: Sufficient power).}
\label{fig:Markov}
\vspace{-0.55cm}
\end{figure}

\vspace{-0.45cm}
\subsection{Power Outage Probability}
The power states of ${U_k}$ are modeled by a Markov chain, as shown in Fig.~\ref{fig:Markov}.
In each time slot, ${U_k}$ can either be in the power sufficient state ${S_1}$ or the power outage state ${S_0}${\footnote{In this paper, we aim to determine if there is a power outage in the energy buffer, which affects the downlink and uplink information transmission. In this paper, two states, i.e., power sufficient and outage states, are used to model the energy buffer\cite{9464720}, where the energy buffer is modeled as two energy levels. In fact, there will be a power outage when the energy is below the second energy level, and vice versa. Thus, two energy levels are enough to model the energy buffer in the proposed system. In addition, simulation of more energy levels\cite{9374454,8105867} is presented in Sect.~V.}}.
The states of ${U_k}$ at the $\left( {q - 1} \right)$\text{-}th, $q$\text{-}th, and $\left( {q + 1} \right)$\text{-}th time slots are illustrated in Fig.~\ref{fig:Markov}.
The transition probabilities between two power states are represented by $P_{uv}^{k,po}$, where $u,v \in \left\{ {0,1} \right\}$ with $0$ corresponding to state ${S_0}$ and $1$ corresponding to state ${S_1}$. Specifically, the transition probabilities from a given state at $q$\text{-}th time slot to the states at $\left( {q + 1} \right)$\text{-}th time slot are expressed as $P_{00}^{k,po}\left( q \right)$, $P_{01}^{k,po}\left( q \right)$, $P_{10}^{k,po}\left( q \right)$, and $P_{11}^{k,po}\left( q \right)$\cite{9464720}.
It is assumed that there are $Q$ consecutive time slots, $q$ denotes the $q$-th time slot $\left( {q = 1,2, \ldots, Q} \right)$, and $C$ is the initial state of the battery.
The residual energy of $q = 1$ and $q = 2$ can be expressed as, respectively
\vspace{-0.3cm}
\begin{small}
\begin{align}
&{B_k}\left( 1 \right) = {E_k}\left( 1 \right) + C - {E_s},\\
&{B_k}\left( 2 \right) = {B_k}\left( 1 \right) + {E_k}\left( 2 \right) - {E_s}.
\vspace{-0.3cm}
\end{align}
\end{small}%
Hence, the residual energy of $q$-th time slot can be summarized as
\vspace{-0.3cm}
\begin{small}
\begin{align}
{B_k}\left( q \right) &= {B_k}\left( {q - 1} \right) + {E_k}\left( q \right) - {E_s}\nonumber\\
& = {B_k}\left( {q - 2} \right) + {E_k}\left( {q - 1} \right) - {E_s} + {E_k}\left( q \right) - {E_s}\nonumber\\
& \ \, \vdots \nonumber\\
& = \sum\limits_{l = 1}^q {{E_k}} \left( l \right) + C - q{E_s},
\vspace{-0.6cm}
\end{align}
\end{small}%
where ${E_k}\left( l \right) \!=\! \eta \left( {1 - \theta } \right)\!{\beta _k}{\left| {\sum\limits_{\ell  = 1}^L {\left(\! {h_{k,\ell ,l}^0 + \sum\limits_{i = 1}^N {{h_{\ell ,i,l}}{g_{k,i,l}}} }\! \right)} } \right|^2}\!\!\!\frac{{{P_{{\rm{AP}}}}}}{L}{T_s}$, ${h_{k,\ell ,l}^0}$, ${{h_{\ell ,i,l}}}$, ${{g_{k,i,l}}}$ denote the time-varying channel magnitudes at time slot $l$ with $l \in \left\{ {1,2, \ldots ,q} \right\}$.

To ensure that no power outage occurs at the $q$-th time slot, ${B_k}\left( q \right)$ should be greater than or equal to $0$. Thus,
we can calculate the transmission probability $P_{11}^{k,po}\left( q \right)$ to ensure that the power is sufficient at the $q$-th time slot (i.e., the cumulative harvested energy of the energy buffer should not be less than the consumption).
$P_{11}^{k,po}\left( q \right)$ is given by
\vspace{-0.20cm}
\begin{small}
\begin{align}
P_{11}^{k,po}\left( q \right) &= \Pr \left( {\sum\limits_{l = 1}^q {{{E_k}} \left( l \right)}  + C \ge q{E_s}} \right)\nonumber\\
 &= \Pr \left( {\sum\limits_{l = 1}^q {\eta \left( {1 - \theta } \right){\beta _k}\frac{{{P_{\text{AP}}}}}{L}{T_s}{\cal Z}_l^2}  + C \ge q{E_s}} \right)\nonumber\\
& = \Pr \left( {\sum\limits_{l = 1}^q {{\cal Z}_l^2}  \ge \frac{{L\left( {q{E_s} - C} \right)}}{{\eta \left( {1 - \theta } \right){\beta _k}{P_{\text{AP}}}{T_s}}}} \right),
\vspace{-0.2cm}
\end{align}
\end{small}%
where $\mathcal{Z}_l = {\sum\limits_{\ell  = 1}^L {\left( {h_{k,\ell,l }^0 + \sum\limits_{i = 1}^N {{h_{\ell ,i,l}}{g_{k,i,l}}} } \right)} } $.

We first derive the probability density function (PDF) of ${\sum\limits_{l = 1}^q {{\mathcal{Z}_l^2}} }$.
Let ${{\mathcal{V}_{\ell il }}} = {{h_{\ell,i,l}}{g_{k ,i,l}}}$, and the $n$-th moment of ${{\mathcal{V}_{\ell il }}}$ can be expressed as shown in (\ref{eq:kth}) \cite{9406837}.
\vspace{-0.25cm}
\begin{small}
\begin{equation}
\label{eq:kth}
{\mu _{{{\mathcal{V}_{\ell il }}}}}\left( n \right) = \lambda_{\ell i} ^{ - n}\frac{{\Gamma \left( {{m_{\ell,i }} + n/2} \right)\Gamma \left( {{m_{k,i}} + n/2} \right)}}{{\Gamma \left( {{m_{\ell,i }}} \right)\Gamma \left( {{m_{k,i}}} \right)}},
\vspace{-0.2cm}
\end{equation}
\end{small}%
where ${\lambda_{\ell i}} = \sqrt {\frac{{{m_{\ell,i}}}}{{{\Omega _{\ell,i }}}}\frac{{{m_{k,i}}}}{{{\Omega _{k,i}}}}}$.

Let ${G_{\ell l} } = \sum\nolimits_{i = 1}^N {{\mathcal{V}_{\ell il}}} $. Using multinomial expansion \cite{5740505}, the $n$-th moment of ${G_{\ell l}}$, i.e., ${\mu _{{G_{\ell l}}}}\left( n \right) \buildrel \Delta \over = {\rm{E}}\left( {G_{\ell l} ^n} \right)$, can be obtained as
\vspace{-0.4cm}
\begin{small}
\begin{align}
\label{eq:Gth}
&{\mu _{{G_{\ell l} }}}\left( n \right) = \sum\limits_{{n_1} = 0}^n {\sum\limits_{{n_2} = 0}^{{n_1}} { \cdots \sum\limits_{{n_{N - 1}} = 0}^{{n_{N - 2}}} {\left( {\begin{array}{*{20}{c}}
n\\
{{n_1}}
\end{array}} \right)} } } \left( {\begin{array}{*{20}{c}}
{{n_1}}\\
{{n_2}}
\end{array}} \right) \cdots \left( {\begin{array}{*{20}{c}}
{{n_{N - 2}}}\\
{{n_{N - 1}}}
\end{array}} \right)\nonumber\\
 &\times {\mu _{{\mathcal{V}_{\ell il }}}}\left( {n - {n_1}} \right){\mu _{{\mathcal{V}_{\ell il }}}}\left( {{n_1} - {n_2}} \right) \cdots {\mu _{{\mathcal{V}_{\ell il }}}}\left( {{n_{N - 1}}} \right).
 \vspace{-0.5cm}
\end{align}
\end{small}%
From (\ref{eq:kth}) and (\ref{eq:Gth}),  the first and second moments of ${G_{\ell l}}$ is obtained as
\vspace{-0.2cm}
\begin{small}
\begin{align}
&{\mu _{{G_{\ell l}}}}\left( 1 \right) = \sum\limits_{i = 1}^N {{\mu _{{{\cal V}_{\ell il}}}}\left( 1 \right)} ,\\
&{\mu _{{G_{\ell l}}}}\left( 2 \right) = \sum\limits_{i = 1}^N {{\mu _{{{\cal V}_{\ell il}}}}\left( 2 \right)}  + 2\sum\limits_{i = 1}^N {{\mu _{{{\cal V}_{\ell il}}}}\left( 1 \right)} \!\!\sum\limits_{i' = i + 1}^N \!\! {\mu _{{{\cal V}_{\ell i'l}}}}\left( 1 \right).
\vspace{-0.3cm}
\end{align}
\end{small}%
Since $h_{k,\ell,l }^0\sim{\rm{Nakagami}}\left( {m_{k,\ell }^0,\Omega _{k,\ell }^0} \right)$, one has
\vspace{-0.2cm}
\begin{small}
\begin{eqnarray}
{\mu _{h_{k,\ell,l }^0}}\left( n \right) = \frac{{\Gamma \left( {m_{k,\ell }^0 + n/2} \right)}}{{\Gamma \left( {m_{k,\ell }^0} \right)}}{\left( {\frac{{m_{k,\ell }^0}}{{\Omega _{k,\ell }^0}}} \right)^{ - n/2}}.
\vspace{-0.3cm}
\end{eqnarray}
\end{small}%
Let $\mathcal{P}_{\ell l} = h_{k,\ell,l }^0 + {G_{\ell l} }$. The $n$-th moment of $\mathcal{P}_{\ell l}$ can be expressed as ${\mu _{\mathcal{P}_{\ell l}}}\left( n \right) \buildrel \Delta \over = {\rm{E}}\left( \mathcal{P}_{\ell l} ^n \right) = {\rm{E}}\left( {{{\left( h_{k,\ell,l }^0 + {G_{\ell l} } \right)}^n}} \right)$, where ${h_{k,\ell,l }^0}$ and ${G_{\ell l} }$ are mutually independent. Using the binomial theorem, ${\mu _{\mathcal{P}_{\ell l}}}\left( n \right)$ is calculated as
\vspace{-0.2cm}
\begin{small}
\begin{align}
{\mu _{\mathcal{P}_{\ell l}}}\left( n \right) &= {\rm{E}}\left[ {\sum\limits_{\varepsilon  = 0}^n {\left( {\begin{array}{*{20}{c}}
n\\
\varepsilon
\end{array}} \right)h_{k,\ell,l }^{0,\varepsilon }G_{\ell l} ^{n - \varepsilon }} } \right]\nonumber\\
 &= \sum\limits_{\varepsilon  = 0}^n {\left( {\begin{array}{*{20}{c}}
n\\
\varepsilon
\end{array}} \right){\mu _{h_{k,\ell,l }^0}}\left( \varepsilon  \right){\mu _{{G_{\ell l} }}}\left( {n - \varepsilon } \right)}.
\vspace{-0.2cm}
\end{align}
\end{small}%
Hence, one has
\vspace{-0.2cm}
\begin{small}
\begin{align}
&{\mu _{\mathcal{P}_{\ell l}}}\left( 1 \right) = {\mu _{h_{k,\ell,l }^0}}\left( 1 \right) + {\mu _{G_{\ell l} }}\left( 1 \right),\\
&{\mu _{\mathcal{P}_{\ell l}}}\left( 2 \right) = {\mu _{h_{k,\ell,l }^0}}\left( 2 \right) + {\mu _{G_{\ell l} }}\left( 2 \right) + 2{\mu _{{h_{k,\ell,l }^0}}}\left( 1 \right){\mu _{G_{\ell l} }}\left( 1 \right).
\vspace{-0.2cm}
\end{align}
\end{small}%
The distribution of ${\mathcal{P}_{\ell l}}$ can be approximated with a Gamma distribution by the moment-matching approach\cite{9138463,9774334}, i.e.,
\vspace{-0.1cm}
\begin{small}
\begin{equation}
{\mathcal{P}_{\ell l}} \sim \text{Gamma}\left( {\chi ,\phi   } \right),
\vspace{-0.2cm}
\end{equation}
\end{small}%
where
\vspace{-0.2cm}
\begin{small}
\begin{eqnarray}
\chi  = \frac{{{{\left[ {{\rm{E}}\left( {\mathcal{P}_{\ell l}} \right)} \right]}^2}}}{{{\rm{Var}}\left( {\mathcal{P}_{\ell l}} \right)}} = \frac{{{{\left[ {{\mu _{\mathcal{P}_{\ell l}}}\left( 1 \right)} \right]}^2}}}{{{\mu _{\mathcal{P}_{\ell l}}}\left( 2 \right) - {{\left[ {{\mu _{\mathcal{P}_{\ell l}}}\left( 1 \right)} \right]}^2}}},\\
\phi  = \frac{{{\rm{E}}\left( {\mathcal{P}_{\ell l}} \right)}}{{{\rm{Var}}\left( {\mathcal{P}_{\ell l}} \right)}} = \frac{{{\mu _{\mathcal{P}_{\ell l}}}\left( 1 \right)}}{{{\mu _{\mathcal{P}_{\ell l}}}\left( 2 \right) - {{\left[ {{\mu _{\mathcal{P}_{\ell l}}}\left( 1 \right)} \right]}^2}}}.
\vspace{-0.2cm}
\end{eqnarray}
\end{small}%

Since ${\cal Z}_l = \sum\limits_{\ell  = 1}^L {{\mathcal{P}_{\ell l}}}$, one has ${\mathcal{Z}_l} \sim \text{Gamma}\left( {L\chi , \phi  } \right)$. The mean and variance of ${\cal Z}_l$ are ${\rm{E}}\left( {\cal Z}_l \right) = L\chi /\phi$ and ${\rm{Var}}\left( {\cal Z}_l \right) = L\chi /{\phi ^2}$, respectively.
Let $\sum\limits_{l = 1}^q {{\cal Z}_l^2}  = \mathcal{T}$. $\mathcal{T}$ follows the non-central chi-square distribution with $q$ degrees of freedom. Therefore, the PDF of $\mathcal{T}$ is expressed as
\vspace{-0.2cm}
\begin{small}
\begin{align}
f_\mathcal{T}\left( x \right) = \frac{1}{{2{\sigma ^2}}}{\left( {\frac{x}{{{s^2}}}} \right)^{\frac{{q - 2}}{4}}}{e^{ - \frac{{{s^2} + x}}{{2{\sigma ^2}}}}}{I_{\frac{q}{2} - 1}}\left( {\frac{s}{{{\sigma ^2}}}\sqrt x } \right),
\vspace{-0.2cm}
\end{align}
\end{small}%
where ${\sigma ^2} = {\rm{Var}}\left( \mathcal{Z}_l \right)$ and ${s} = \sqrt {q{{\left( {{\rm{E}}\left( \mathcal{Z}_l \right)} \right)}^2}}  $.
Thus, ${P^{k,po}_{11}}\left( q \right)$ can be calculated as
\vspace{-0.15cm}
\begin{small}
\begin{align}
\label{eq:P11}
&{P^{k,po}_{11}}\left( q \right) = \int_{\frac{{L\left( {q{E_s} - C} \right)}}{{\eta \left( {1 - \theta } \right){\beta _k }{P_{\text{AP}}}{T_s}}}}^\infty  {f_\mathcal{T}\left( x \right)} dx\nonumber\\
 &\!=\! \sum\limits_{\varsigma  = 0}^\infty \!\!\! {\sum\limits_{p = 0}^{^{\frac{q}{2} - 1 + \varsigma}}\!\!\!\! {\frac{{{s^{2\varsigma }}{e^{ - \frac{{{s^2}}}{{2{\sigma ^2}}}}}}}{{{{\left( {2{\sigma ^2}} \right)}^{\varsigma  + p}}\varsigma !}}} } {e^{ - \frac{{L\left( {q{E_s} - C} \right)}}{{2{\sigma ^2}\eta \left( {1 - \theta } \right){\beta _k }{P_{\text{AP}}}{T_s}}}}}\frac{{{{\left(\! {\frac{{L\left( {q{E_s} - C} \right)}}{{\eta \left( {1 - \theta } \right){\beta _k }{P_{\text{AP}}}{T_s}}}} \!\right)}^p}}}{{p!}}.
 \vspace{-0.2cm}
\end{align}
\end{small}%
The detailed derivation of (\ref{eq:P11}) is given in Appendix A. When ${\beta _k} = 0$, one has $\mathop {\lim }\limits_{{\beta _k} = 0} \frac{{L\left( {q{E_s} - C} \right)}}{{\eta \left( {1 - \theta } \right){\beta _k}{P_{{\rm{AP}}}}{T_s}}} = \infty $. Using $\mathop {\lim }\limits_{x \to \infty } {e^{ - x}}x = 0$, (\ref{eq:P11}) can be expressed as $P_{11}^{k,po}\left( q \right) = 0$. To ensure $P_{11}^{k,po}\left( q \right) > 0$, ${\beta _k}$ should be greater than zero.

It is noted that $q \ge 2$ in (\ref{eq:P11}) satisfies $\left( {\frac{q}{2} - 1 + \varsigma } \right) \ge 0$. For the special case of $q = 1$, one has
\begin{small}
\begin{align}
 \vspace{-0.2cm}
\label{eq:P112}
P_{11}^{k,po}\left( {q = 1} \right) = \Pr \left( {{{\mathcal{Z}_l^2}} \ge \frac{{L\left( {q{E_s} - C} \right)}}{{\eta \left( {1 - \theta } \right){\beta _k}{P_{AP}}{T_s}}}} \right).
 \vspace{-0.2cm}
\end{align}
\end{small}%
Since ${\mathcal{Z}_l} \sim \text{Gamma}\left( {L\chi ,  \phi } \right)$, the CDF of $\mathcal{Z}_l$ can be expressed as
\vspace{-0.2cm}
\begin{small}
\begin{equation}
{F_{{\mathcal{Z}_l}}}\left( x \right) = \frac{1}{{\Gamma \left( {L\chi} \right)}}\gamma \left( {L\chi,\phi x} \right).
 \vspace{-0.2cm}
\end{equation}
\end{small}%
For $Y = {X^2}$, the CDF of $Y$ can be calculated as ${F_Y}\left( y \right) = {F_X}\left( {\sqrt {{y}} } \right)$. The CDF of $\mathcal{Z}_l^2$ is computed as
\vspace{-0.2cm}
\begin{small}
\begin{equation}
\label{eq:CDFtP1}
{F_{\mathcal{Z}_l^2}}\left( x \right) = \frac{1}{{\Gamma \left( {L\chi} \right)}}\gamma \left( {L\chi,\phi {x^{\frac{1}{2}}}} \right).
 \vspace{-0.2cm}
\end{equation}
\end{small}%
The PDF of $\mathcal{Z}_l^2$ is obtained as
\vspace{-0.2cm}
\begin{small}
\begin{align}
\label{eq:PDFtP1}
{f_{{\cal Z}_l^2}}\left( x \right) = \frac{{{\phi ^{L\chi }}{e^{ - \phi {x^{\frac{1}{2}}}}}{x^{\frac{{L\chi  - 2}}{2}}}}}{{2\left( {L\chi  - 1} \right)!}}.
 \vspace{-0.3cm}
\end{align}
\end{small}%

\begin{figure}
\center
\includegraphics[width=3.58in,height=2.25in]{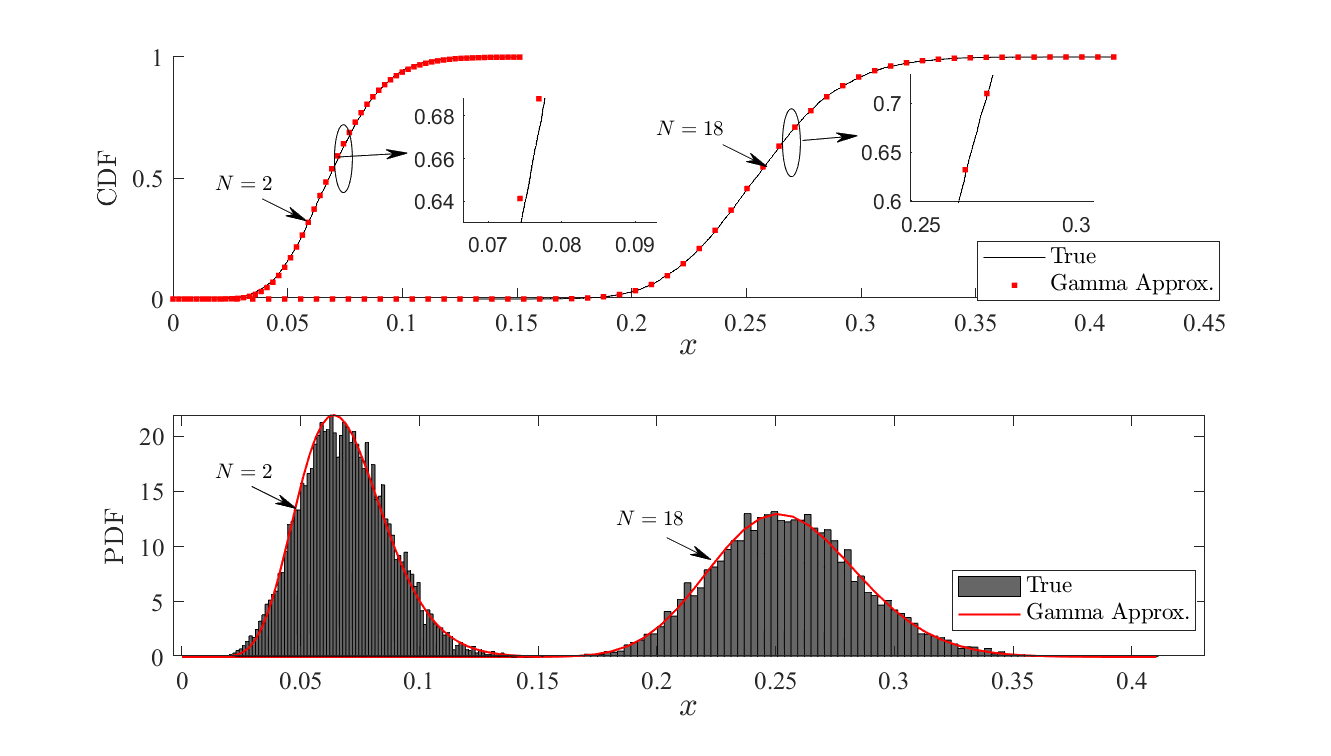}
\vspace{-0.25cm}
\captionsetup{font={footnotesize}}
\caption{The CDF and PDF of ${\mathcal{P}_{\ell l}}$ for the true distribution and the approximate Gamma distribution.}
\label{fig:gamma_approx}
\vspace{-0.95cm}
\end{figure}

Thus, (\ref{eq:P112}) can be calculated as
\vspace{-0.2cm}
\begin{small}
\begin{align}
{P^{k,po}_{11}}\!\left(\! {q = 1} \!\right) \!=\! 1 \!-\! \frac{1}{{\Gamma \left( {L\chi} \right)}}\gamma\! \left( \!\!{L\chi,\phi {{\left(\! \frac{{L\left( \!{q{E_s} - C} \!\right)}}{{\eta \left( \!{1 - \theta } \!\right){\beta _k }{P_{\text{AP}}}{T_s}}} \!\right)}^{\!\!\frac{1}{2}}}} \!\right)\!.
 \vspace{-0.2cm}
\end{align}
\end{small}%

According to the Markov model presented in Fig.~\ref{fig:Markov}, the power outage probability $P^{k,po}$ of ${U_k}$ over $Q$ time slots is calculated as
\vspace{-0.3cm}
\begin{small}
\begin{align}
\label{eq:Pp0}
P^{k,po} = 1 - \prod\limits_{q = 1}^Q {{P^{k,po}_{11}}\left( q \right)}.
 \vspace{-0.5cm}
\end{align}
\end{small}%

{{\em Remark~2:} For the buffer-less system, ${{P^{k,po}_{11}}\left( q \right)}$ degenerates to ${{P^{k,po}_{11}}\left( {q = 1} \right)}$ at each time slot. The power sufficient probability of a buffer-less system is $\prod\limits_{q = 1}^Q {P_{11}^{k,po}\left( {1} \right)}$.
Hence, the overall power outage $P_{bl}^{k,po}$ of a buffer-less system over $Q$ time slots can be expressed as $P_{bl}^{k,po} = 1 - {\left( {P_{11}^{k,po}\left( {1} \right)} \right)^Q}$.

{{\em Remark~3:} Fig.~\ref{fig:gamma_approx} presents the cumulative distribution function (CDF) and PDF of ${\mathcal{P}_{\ell l}}$ for the true distribution and the approximate Gamma distribution. The true distribution is generated through Monte Carlo simulations. It can be observed that the approximate CDF and PDF are consistent with true CDF and PDF, respectively, which validates the correctness of our analysis. The fitted Gamma distribution becomes more accurate with a larger number of STAR-RIS elements\cite{9808307}.
In addition, from the simulation and theory analysis results in Sect.~V, it can be also seen that the simulated curves match with the theoretical ones.}

{{\em Proposition~3.1:} The minimum power outage probability can be achieved when the initial state of the battery satisfies the overall power consumption over $Q$ time slots, i.e., $C = Q{E_s}$.

{{\em Proof:} We take the derivative of ${P_{11}^{k,po}\left( q \right)}$. One has
\vspace{-0.15cm}
\begin{small}
\begin{align}
&\frac{{\partial P_{11}^{k,po}\left( q \right)}}{{\partial C}} = \sum\limits_{\varsigma  = 0}^\infty   \frac{{{s^{2\varsigma }}{e^{ - \frac{{{s^2}}}{{2{\sigma ^2}}}}}}}{{{{\left( {2{\sigma ^2}} \right)}^\varsigma }\varsigma !}}\frac{{{e^{ - \frac{{L\left( {q{E_s} - C} \right)}}{{2{\sigma ^2}\eta \left( {1 - \theta } \right){\beta _k}{P_{{\rm{AP}}}}{T_s}}}}}L}}{{2{\sigma ^2}\eta \left( {1 - \theta } \right){\beta _k}{P_{{\rm{AP}}}}{T_s}}}\nonumber\\
& \times \!\left( \!{\sum\limits_{p = 0}^{\frac{q}{2} - 1 + \varsigma } {\frac{{{{\left( {\frac{{L\left( {q{E_s} - C} \right)}}{{\eta \left( {1 - \theta } \right){\beta _k}{P_{{\rm{AP}}}}{T_s}}}} \right)}^p}}}{{p!}} - \sum\limits_{p = 1}^{\frac{q}{2} - 1 + \varsigma } {\frac{{{{\left( {\frac{{L\left( {q{E_s} - C} \right)}}{{\eta \left( {1 - \theta } \right){\beta _k}{P_{{\rm{AP}}}}{T_s}}}} \right)}^{p - 1}}}}{{\left( {p - 1} \right)!}}} } } \!\right)\nonumber\\
&\! = \!\sum\limits_{\varsigma  = 0}^\infty   \frac{{{s^{2\varsigma }}{e^{ - \frac{{{s^2}}}{{2{\sigma ^2}}}}}}}{{{{\left( {2{\sigma ^2}} \right)}^\varsigma }\varsigma !}}\frac{{{e^{ - \frac{{L\left( {q{E_s} - C} \right)}}{{2{\sigma ^2}\eta \left( {1 - \theta } \right){\beta _k}{P_{{\rm{AP}}}}{T_s}}}}}L}}{{2{\sigma ^2}\eta \left( {1 - \theta } \right){\beta _k}{P_{{\rm{AP}}}}{T_s}}}\frac{{{{\left( \!{\frac{{L\left( {q{E_s} - C} \right)}}{{\eta \left( {1 - \theta } \right){\beta _k}{P_{{\rm{AP}}}}{T_s}}}} \!\right)}^{\frac{q}{2} - 1 + \varsigma }}}}{{\left( {\frac{q}{2} - 1 + \varsigma } \right)!}}.
\vspace{-0.4cm}
\end{align}
\end{small}%
By assigning $\frac{{\partial P_{11}^{k,po}\left( q \right)}}{{\partial C}} = 0$, one has $C = q{E_s}$. Considering $Q$ time slots, we have that if $C = Q{E_s}$, the power outage probability is minimized over $Q$ time slots.}}

{{\em Proposition~3.2:} The average residual energy after $Q$ time slots is given by ${C_R} = \eta \left( {1 - \theta } \right){\beta _k}{P_{{\rm{AP}}}}{T_s}{\rm{Q}}\frac{{\chi \left( {L\chi  + 1} \right)}}{{{\phi ^2}}} + C - Q{E_s}$. Only if $\eta \left( {1 - \theta } \right){\beta _k}{P_{{\rm{AP}}}}{T_s}\frac{{\chi \left( {L\chi  + 1} \right)}}{{{\phi ^2}}} \ge {E_s}$, a sustainable wireless node is possible. Otherwise, the system will gradually deteriorate to a buffer-less system.

{{\em Proof:} We perform the expectation operation on the residual energy after $Q$ time slots. One has
\vspace{-0.2cm}
\begin{small}
\begin{align}
{C_R} &= {\rm{E}}\left( {\sum\limits_{l = 1}^Q {{E_k}}  + C - Q{E_s}} \right)\nonumber\\
 &= {\rm{E}}\left( {\sum\limits_{l = 1}^Q {\eta \left( {1 - \theta } \right){\beta _k}\frac{{{P_{{\rm{AP}}}}}}{L}{T_s}Z_l^2}  + C - Q{E_s}} \right)\nonumber\\
 &= \eta \left( {1 - \theta } \right){\beta _k}\frac{{{P_{{\rm{AP}}}}}}{L}{T_s}{\rm{QE}}\left( {Z_l^2} \right) + C - Q{E_s}\nonumber\\
 &= \eta \left( {1 - \theta } \right){\beta _k}{P_{{\rm{AP}}}}{T_s}{\rm{Q}}\frac{{\chi \left( {L\chi  + 1} \right)}}{{{\phi ^2}}} + C - Q{E_s}.
 \vspace{-0.2cm}
\end{align}
\end{small}%

To provide more useful insights, the asymptotic ${P^{k,po}_{11}}\left( q \right)$ against ${{P_{{\rm{AP}}}}}$ is investigated.

{\em Lemma~1:} An asymptotic expression for ${P^{k,po}_{11}}\left( q \right)$ at high ${{P_{{\rm{AP}}}}}$ values can be expressed as
\begin{align}
P_{11}^{k,po}\left( q \right) = \sum\limits_{\varsigma  = 0}^\infty   \frac{{{s^{2\varsigma }}{e^{ - \frac{{{s^2}}}{{2{\sigma ^2}}}}}}}{{{{\left( {2{\sigma ^2}} \right)}^\varsigma }\varsigma !}}.
\end{align}

{\em Proof:} At high $P_\text{AP}$ values, one can get $\frac{{L\left( {q{E_s} - C} \right)}}{{\eta \left( {1 - \theta } \right){\beta _k}{P_{{\rm{AP}}}}{T_s}}} \to 0$. Thus, in (\ref{eq:P11}), the inner summation makes sense only for ${p = 0}$. The inner summation becomes $\frac{{{s^{2\varsigma }}{e^{ - \frac{{{s^2}}}{{2{\sigma ^2}}}}}}}{{{{\left( {2{\sigma ^2}} \right)}^\varsigma }\varsigma !}}$.

{\em Remark~4:} Using the asymptotic ${P^{k,po}_{11}}\left( q \right)$, the asymptotic power outage probability $P^{k,po}$ can be obtained as ${P^{k,po}} = 1 - \prod\limits_{q = 1}^Q {\sum\limits_{\varsigma  = 0}^\infty   \frac{{{s^{2\varsigma }}{e^{ - \frac{{{s^2}}}{{2{\sigma ^2}}}}}}}{{{{\left( {2{\sigma ^2}} \right)}^\varsigma }\varsigma !}}} $. The diversity order of ${P^{k,po}}$ can be calculated as $\mathop {\lim }\limits_{{P_{{\rm{AP}}}} \to \infty }  - \frac{{\log \left( {{P^{k,po}}\left( {{P_{{\rm{AP}}}}} \right)} \right)}}{{\log \left( {{P_{{\rm{AP}}}}} \right)}} = 0$, since $P^{k,po}$  is independent with ${{P_{{\rm{AP}}}}}$.

\subsection{Downlink Information Outage Probability and Sum Throughput}
The primary goal of energy collection is to provide the energy for exchanging information.
When only part of the received power is available for information communication, it is crucial to assess the probability of communication outages.
According to the principle of NOMA, the user with better channel conditions performs SIC decoding{\footnote{In this paper, perfect SIC is considered, which brings an upper bound for the information outage probability of the proposed system. Due to unavoidable error propagation and quantization error in practical hardware implementation, imperfect SIC is more realistic\cite{10144629,9984815}. Hence, simulation results are presented in  Sect.~V to investigate the impact of imperfect SIC on the system performance.}}, while the other user decodes its signal directly\cite{9774334}. In this paper, based on the path loss\cite{9774334,9722712,9856598}{\footnote{In this paper, it is assumed that ${d_{S\text{-}R,{U_t}}} < {d_{S\text{-}R,{U_r}}}$. This causes more severe path loss for ${U_r}$ compared to ${U_t}$. To ensure user fairness, more transmit power is allocated to ${U_r}$, i.e., ${\alpha _r} > {\alpha _t}$.}}, it is assumed that ${U_t}$ experiences better channel conditions, and there is a fixed decoding order{\footnote{In this paper, we can adopt other decoding orders, such as\cite{9188014}. The simulation result is presented in Sect.~V to show the impact of decoding order on information outage probability performance.}} $\left( {{U_r},{U_t}} \right)$ at ${U_t}$.

According to (\ref{eq:yk}) and the decoding order $\left( {{U_r},{U_t}} \right)$, ${U_t}$ first decodes the signal of ${U_r}$ before decoding its
own signal. The signal-to-interference plus noise ratio (SINR) at $U_t$ is given by
\vspace{-0.3cm}
\begin{small}
\begin{align}
\label{eq:rds}
{\gamma _s^d} = \frac{{\theta {\beta _t}{{\left| {\sum\limits_{\ell  = 1}^L {\left( {h_{t,\ell }^0 + \sum\limits_{i = 1}^N {{h_{\ell ,i}}{g_{t,i}}} } \right)} } \right|}^2}{\alpha _r}{P_{\text{AP}}}}}{{\theta {\beta _t}{{\left| {\sum\limits_{\ell  = 1}^L {\left( {h_{t,\ell }^0 + \sum\limits_{i = 1}^N {{h_{\ell ,i}}{g_{t,i}}} } \right)} } \right|}^2}{\alpha _t}{P_{\text{AP}}} + L{N_0}}}.
\vspace{-0.2cm}
\end{align}
\end{small}%
Then, ${U_t}$ performs SIC to remove the interference from the signal of ${U_r}$ and decodes its signal. The SNR at $U_t$ can be expressed as
\vspace{-0.25cm}
\begin{small}
\begin{align}
\label{eq:rdt}
{\gamma _t^d} = \frac{{\theta {\beta _t}{{\left| {\sum\limits_{\ell  = 1}^L {\left( {h_{t,\ell }^0 + \sum\limits_{i = 1}^N {{h_{\ell ,i}}{g_{t,i}}} } \right)} } \right|}^2}{\alpha _t}{P_{\mathrm{AP}}}}}{{L{N_0}}}.
\vspace{-0.2cm}
\end{align}
\end{small}%

\begin{figure}
\center
\includegraphics[width=2.4in,height=0.8in]{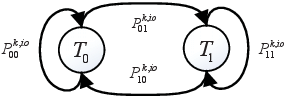}
\vspace{-0.10cm}
\captionsetup{font={footnotesize}}
\caption{The Markov chain model of information transmission states (${T_0}$: Outage, and ${T_1}$: Successful transmission).}
\label{fig:Markov2}
\end{figure}
Meanwhile, ${U_r}$ treats the signal of ${U_t}$ as interference and decodes its own signal directly. The SINR at $U_r$ is obtained as
\vspace{-0.3cm}
\begin{small}
\begin{align}
\label{eq:rdr}
{\gamma _r^d} = \frac{{\theta {\beta _r}{{\left| {\sum\limits_{\ell  = 1}^L {\left( {h_{r,\ell }^0 + \sum\limits_{i = 1}^N {{h_{\ell ,i}}{g_{r,i}}} } \right)} } \right|}^2}{\alpha _r}{P_{\mathrm{AP}}}}}{{\theta {\beta _r}{{\left| {\sum\limits_{\ell  = 1}^L {\left( {h_{r,\ell }^0 + \sum\limits_{i = 1}^N {{h_{\ell ,i}}{g_{r,i}}} } \right)} } \right|}^2}{\alpha _t}{P_{\mathrm{AP}}} + L{N_0}}}.
\vspace{-0.2cm}
\end{align}
\end{small}%
Similar to the power outage analysis in section~III-A, the information transmission states can also be modeled with a Markov chain, as illustrated in Fig.~\ref{fig:Markov2}. The Markov
chain can achieve a steady state, defined as
\begin{small}
\begin{align}
\left\{ {\begin{array}{*{20}{c}}
{{\pi _{0,d}^k}P_{00,d}^{k,io} + {\pi _{1,d}^k}P_{10,d}^{k,io} = {\pi _{0,d}^k}},\\
{{\pi _{0,d}^k}P_{01,d}^{k,io} + {\pi _{1,d}^k}P_{11,d}^{k,io} = {\pi _{1,d}^k}},
\end{array}} \right.
\vspace{-0.3cm}
\end{align}
\end{small}%
where ${\pi _{0,d}^k}$ and ${\pi _{1,d}^k}$ are the steady state probabilities for two
states ${T_0}$ and ${T_1}$, ${\pi _{0,d}^k} + {\pi _{1,d}^k} = 1$, $P_{00,d}^{k,io} + P_{01,d}^{k,io} = 1$, and $P_{10,d}^{k,io} + P_{11,d}^{k,io} = 1$. Thus, one has
\begin{small}
\begin{align}
\vspace{-0.4cm}
\left\{ {\begin{array}{*{20}{c}}
{{\pi _{0,d}^k} = \frac{{P_{10,d}^{k,io}}}{{P_{01,d}^{k,io} + P_{10,d}^{k,io}}}},\\
{{\pi _{1,d}^k} = \frac{{P_{01,d}^{k,io}}}{{P_{01,d}^{k,io} + P_{10,d}^{k,io}}}}.
\end{array}} \right.
\end{align}
\end{small}%
\subsubsection{Downlink Information Outage Probability of ${U_r}$}
The transition probability $P_{10,d}^{r,io}$ denotes the outage probability of ${U_r}$, which is calculated as
\begin{small}
\begin{align}
\label{eq:P10r}
&P_{10,d}^{r,io} = \Pr \left( {{\gamma _r^d} < {\gamma _{th}}} \right)\nonumber\\
 &= \Pr \left( {\frac{{\theta {\beta _r}{{\left| {\sum\limits_{\ell  = 1}^L {\left( {h_{r,\ell }^0 + \sum\limits_{i = 1}^N {{h_{\ell ,i}}{g_{r,i}}} } \right)} } \right|}^2}{\alpha _r}{P_{\mathrm{AP} }}}}{{\theta {\beta _r}{{\left| {\sum\limits_{\ell  = 1}^L {\left( {h_{r,\ell }^0 + \sum\limits_{i = 1}^N {{h_{\ell ,i}}{g_{r,i}}} } \right)} } \right|}^2}{\alpha _t}{P_{\mathrm{AP} }} + L{N_0}}} < {\gamma _{th}}} \right)\nonumber\\
& = \Pr \left( {{\mathcal{Z}_l^2} < \frac{{L{N_0}{\gamma _{th}}}}{{\theta {\beta _r}{P_{\mathrm{AP} }}\left( {{\alpha _r} - {\alpha _t}{\gamma _{th}}} \right)}}} \right)\nonumber\\
& = \frac{1}{{\Gamma \left( L\chi  \right)}}\gamma \left( {L\chi ,\phi {{\left( {\frac{{L{N_0}{\gamma _{th}}}}{{\theta {\beta _r}{P_{\mathrm{AP} }}\left( {{\alpha _r} - {\alpha _t}{\gamma _{th}}} \right)}}} \right)}^{\frac{1}{2}}}} \right),
\vspace{-0.25cm}
\end{align}
\end{small}%
where ${\gamma _{th}} = {2^R} - 1$ and $R$ is the target rate. $P_{01,d}^{r,io} = \Pr \left( {{\gamma _r^d} \ge {\gamma _{th}}} \right) = 1 - P_{10,d}^{r,io}$. Hence, the steady-state probability $\pi _{1,d}^r = P_{01,d}^{r,io} = 1 - P_{10,d}^{r,io}$.
After $Q$ time slots, the overall information outage probability of ${U_r}$ can be calculated as
\vspace{-0.2cm}
\begin{small}
\begin{align}
\label{eq:Pr10}
{P_d^{r,io}} = 1 - \prod\limits_{q = 1}^Q {\pi _{1,d}^r}  = 1 - {\left( {\pi _{1,d}^r} \right)^Q}.
\vspace{-0.5cm}
\end{align}
\end{small}%
\subsubsection{Downlink Information Outage Probability of ${U_t}$}
Similarly, the $P_{10,d}^{t,io}$ of ${U_t}$ can be expressed as
\vspace{-0.2cm}
\begin{small}
\begin{align}
\label{eq:P10t}
P_{10,d}^{t,io} &= 1 - \Pr \left( {{\gamma _s^d} > {\gamma _{th}},{\gamma _t^d} > {\gamma _{th}}} \right)\nonumber\\
& = 1 - \Pr \left( \!\!\!\begin{array}{l}
\frac{{\theta {\beta _t}{{\left| {\sum\limits_{\ell  = 1}^L {\left( {h_{t,\ell }^0 + \sum\limits_{i = 1}^N {{h_{\ell ,i}}{g_{t,i}}} } \right)} } \right|}^2}{\alpha _r}{P_{\mathrm{AP}}}}}{{\theta {\beta _t}{{\left| {\sum\limits_{\ell  = 1}^L {\left( {h_{t,\ell }^0 + \sum\limits_{i = 1}^N {{h_{\ell ,i}}{g_{t,i}}} } \right)} } \right|}^2}{\alpha _t}{P_{\mathrm{AP}}} + L{N_0}}} > {\gamma _{th}},\\
\frac{{\theta {\beta _t}{{\left| {\sum\limits_{\ell  = 1}^L {\left( {h_{t,\ell }^0 + \sum\limits_{i = 1}^N {{h_{\ell ,i}}{g_{t,i}}} } \right)} } \right|}^2}{\alpha _t}{P_{\mathrm{AP}}}}}{{L{N_0}}} > {\gamma _{th}}
\end{array}\!\!\!\! \right)\nonumber\\
& = 1 - \Pr \left( {\begin{array}{*{20}{l}}
{{\mathcal{Z}_l^2} > \frac{{{\gamma _{th}}L{N_0}}}{{\theta {\beta _t}{P_{\mathrm{AP}}}\left( {{\alpha _r} - {\alpha _t}{\gamma _{th}}} \right)}},}\\
{{\mathcal{Z}_l^2} > \frac{{{\gamma _{th}}L{N_0}}}{{\theta {\beta _t}{\alpha _t}{P_{\mathrm{AP}}}}}}
\end{array}} \right)\nonumber\\
& = \frac{1}{{\Gamma \left( L\chi  \right)}}\gamma \left( {L\chi ,\phi {{\left( {{{\hat \gamma }_t}} \right)}^{\frac{1}{2}}}} \right),
\vspace{-0.25cm}
\end{align}
\end{small}%
where ${{\hat \gamma }_t} = \max \left( {\frac{{{\gamma _{th}}L{N_0}}}{{\theta {\beta _t}{P_{\mathrm{AP}}}\left( {{\alpha _r} - {\alpha _t}{\gamma _{th}}} \right)}},\frac{{{\gamma _{th}}L{N_0}}}{{\theta {\beta _t}{\alpha _t}{P_{\mathrm{AP}}}}}} \right)$. The steady-state probability $\pi _{1,d}^t = 1 - P_{10,d}^{t,io}$.
After $Q$ time slots, the overall information outage probability of ${U_t}$ can be calculated as
\begin{small}
\begin{align}
\vspace{-0.35cm}
&{P_d^{t,io}} = 1 - \prod\limits_{q = 1}^Q {\pi _{1,d}^t}  = 1 - {\left( {\pi _{1,d}^t} \right)^Q}.
\vspace{-0.2cm}
\end{align}
\end{small}%
When ${\beta _k} = 0$, one has $\gamma \left( {L\chi ,\infty } \right)$ in both (\ref{eq:P10r}) and (\ref{eq:P10t}). Since $\gamma \left( {m,x} \right) = \left( {m - 1} \right)!\left( {1 - {e^{ - x}}\sum\limits_{n = 0}^{m - 1} {\frac{{{x^n}}}{{n!}}} } \right)$, using $\mathop {\lim }\limits_{x \to \infty } {e^{ - x}}x = 0$, $\gamma \left( {L\chi ,\infty } \right) = \left( {L\chi  - 1} \right)!$. Hence $P_{10,d}^{r,io} = 1$ and $P_{10,d}^{t,io} = 1$. To prevent the user from a complete information outage, ${\beta _{k}}$ should be greater than zero.

{\em Lemma~2:} The asymptotic expressions for $P_{10,d}^{r,io}$ and $P_{10,d}^{t,io}$
at high ${{P_{{\rm{AP}}}}}$ values can be expressed as
\begin{align}
\label{eq:P10_00}
P_{10,d}^{r,io} = P_{10,d}^{t,io} = 0.
\end{align}

{\em Proof:} At high $P_\text{AP}$ values, one can get $\frac{{L{N_0}{\gamma _{th}}}}{{\theta {\beta _r}{P_{{\rm{AP}}}}\left( {{\alpha _r} - {\alpha _t}{\gamma _{th}}} \right)}} \to 0$ and ${{\hat \gamma }_t} \to 0$. Since $\gamma \left( {L\chi ,0} \right) = 0$, (\ref{eq:P10_00}) can be obtained.

The sum throughput of downlink transmission can be expressed as
\vspace{-0.15cm}
\begin{small}
\begin{eqnarray}
\Psi_d   = R{T_s}\left( {1 - {P_d^{r,io}}} \right) + R{T_s}\left( {1 - {P_d^{t,io}}} \right).
\vspace{-0.5cm}
\end{eqnarray}
\end{small}%

\vspace{-0.35cm}
\subsection{Uplink Information Outage Probability and Sum Throughput}
Different from downlink transmission, the received signal strength of ${U_t}$ may be higher due to the shorter distance from the AP, resulting in a decoding order $\left( {{U_t},{U_r}} \right)$ at the AP.
According to (\ref{eq:yAP}), the SINR at the AP is given by
\vspace{-0.25cm}
\begin{small}
\begin{align}
\gamma _s^u = \frac{{{P_t}{\beta _t}{{\left| {\sum\limits_{\ell  = 1}^L {\left( {h_{t,\ell }^0 + \sum\limits_{i = 1}^N {{h_{\ell ,i}}{g_{t,i}}} } \right)} } \right|}^2}}}{{{P_r}{\beta _r}{{\left| {\sum\limits_{\ell  = 1}^L {\left( {h_{r,\ell }^0 + \sum\limits_{i = 1}^N {{h_{\ell ,i}}{g_{r,i}}} } \right)} } \right|}^2} + L{N_0}}}.
\vspace{-0.25cm}
\end{align}
\end{small}%
After performing successful SIC, AP then decodes ${U_r}$. The received SNR for decoding ${U_r}$ is given by
\vspace{-0.25cm}
\begin{small}
\begin{align}
\gamma _r^u = \frac{{{P_r}{\beta _r}{{\left| {\sum\limits_{\ell  = 1}^L {\left( {h_{r,\ell }^0 + \sum\limits_{i = 1}^N {{h_{\ell ,i}}{g_{r,i}}} } \right)} } \right|}^2}}}{{L{N_0}}}.
\vspace{-0.25cm}
\end{align}
\end{small}%
\subsubsection{Uplink Information Outage Probability of ${U_t}$}
Similar to (\ref{eq:P10r}), one has
\vspace{-0.25cm}
\begin{small}
\begin{align}
\label{eq:P10u}
&P_{10,u}^{t,io} = \Pr \left( {\gamma _s^u < {\gamma _{th}}} \right)\nonumber\\
& \approx 1 - \sum\limits_{m = 0}^{L\chi  - 1}  \frac{{{\phi ^{L\chi  + m}}}}{{2m!\left( {L\chi  - 1} \right)!}}\sum\limits_{w = 1}^W {{\psi _w}} {e^{ - \phi {{\left( {\frac{{{P_r}{\beta _r}{e^{{u_w}}}{\gamma _{th}} + L{N_0}{\gamma _{th}}}}{{{P_t}{\beta _t}}}} \right)}^{\frac{1}{2}}}}}\nonumber\\
& \!\times \!\!{\left(\!\! {\frac{{{P_r}{\beta _r}{e^{{u_w}}}{\gamma _{th}} \!+\! L{N_0}{\gamma _{th}}}}{{{P_t}{\beta _t}}}} \!\!\right)^{\!\!\frac{m}{2}}}\!\!\!\!{e^{ - \phi {e^{\frac{{{u_w}}}{2}}}}}\!\!{e^{{u_w}\frac{{L\chi  - 2}}{2}}}\!\!{e^{{u_w} + u_w^2}} \!-\! {O_W}.\!\!
\vspace{-0.25cm}
\end{align}
\end{small}%
The detailed derivation of (\ref{eq:P10u}) is given in Appendix B.

\subsubsection{Uplink Information Outage Probability of ${U_r}$}
\begin{small}
\begin{align}
\label{eq:P10ur}
&P_{10,u}^{r,io} = 1 - \Pr \left( {\gamma _s^u > {\gamma _{th}},\gamma _r^u > {\gamma _{th}}} \right)\nonumber \\
&\approx {F_{{\cal Z}_{l,r}^2}}\left( B \right) + {F_{{\cal Z}_{l,t}^2}}\left( A \right) + \sum\limits_{m = 0}^{L\chi  - 1}  \frac{{{\phi ^{L\chi  + m}}}}{{2m!\left( {L\chi  - 1} \right)!}}\sum\limits_{w = 1}^W {{\psi _w}}\nonumber \\
& \times {e^{ - \phi {{\left( {\frac{{{P_t}{\beta _t}\left( {{e^{{u_w}}} + A} \right) - {\gamma _{th}}L{N_0}}}{{{\gamma _{th}}{P_r}{\beta _r}}}} \right)}^{\frac{1}{2}}}}}{\left( {\frac{{{P_t}{\beta _t}\left( {{e^{{u_w}}} + A} \right) - {\gamma _{th}}L{N_0}}}{{{\gamma _{th}}{P_r}{\beta _r}}}} \right)^{\frac{m}{2}}}\nonumber\\
& \times {e^{ - \phi {{\left( {{e^{{u_w}}} + A} \right)}^{\frac{1}{2}}}}}{\left( {{e^{{u_w}}} + A} \right)^{\frac{{L\chi  - 2}}{2}}}{e^{{u_w} + u_w^2}} + {O_W},
\vspace{-0.25cm}
\end{align}
\end{small}%
where $A = \frac{{{\gamma _{th}}L{N_0}}}{{{P_t}{\beta _t}}}$ and $B = \frac{{{\gamma _{th}}L{N_0}}}{{{P_r}{\beta _r}}}$.
The detailed derivation of (\ref{eq:P10ur}) is given in Appendix B.

For both users, the overall information outage probability is given by
\vspace{-0.5cm}
\begin{small}
\begin{align}
P_u^{^{k,io}} = 1 - \prod\limits_{q = 1}^Q {\pi _{1,u}^k}  = 1 - {\left( {\pi _{1,u}^k} \right)^Q},
\vspace{-0.6cm}
\end{align}
\end{small}%
where $\pi _{1,u}^k = 1 - P_{10,u}^{k,io}$.

The sum throughput of uplink transmission can be expressed as
\vspace{-0.25cm}
\begin{small}
\begin{align}
\label{eq:Th}
{\Psi _u} = R{T_s}\left( {1 - P_u^{r,io}} \right) + R{T_s}\left( {1 - P_u^{t,io}} \right).
\vspace{-0.4cm}
\end{align}
\end{small}%

\vspace{-0.55cm}
\subsection{Joint Outage Probability}
To combine the information outage probability and power outage probability{\footnote{If there is a power outage, a delay occurs, which will be explored in the future.}}, the downlink and uplink joint outage probabilities can be obtained.
For ${U_k}$, the downlink and uplink joint outage probabilities are expressed as, respectively
\vspace{-0.15cm}
\begin{small}
\begin{align}
&P_{{\rm{out}}}^{k,d} = 1 - \left( {1 - {P^{k,po}}} \right)\left( {1 - P_d^{k,io}} \right),\\
&P_{{\rm{out}}}^{k,u} = 1 - \left( {1 - {P^{k,po}}} \right)\left( {1 - P_u^{k,io}} \right).
\end{align}
\end{small}%
Both the downlink and uplink joint outage probabilities are associated with power outage probability. Hence, if a power outage occurs in the downlink joint outage probability, then the uplink joint outage probability occurs.

\section{Resource Allocation}
In this section, we propose a low-complexity power allocation scheme for solving the uplink sum throughput maximization problem with the uplink joint outage probability and JFI constraint.
According to (\ref{eq:rdt}) and (\ref{eq:rdr}), the downlink users' achievable rate is given by
${R_k} = {\log _2}\left( {1 + \gamma _k^d} \right)$. The JFI can be expressed as $JFI = \frac{{{{\left( {{R_t} + {R_r}} \right)}^2}}}{{2\left( {R_t^2 + R_r^2} \right)}}$\cite{10102660}.

\subsection{Problem Formulation}
Our goal is to maximize the uplink sum throughput in the proposed system via jointly optimizing the energy-splitting ratios of the STAR-RIS and the transmit power allocation factors, subject to the uplink joint outage probability and JFI requirements of ${U_r}$ and ${U_t}$.
The uplink sum throughput optimization problem can be formulated as\footnote{Different from (\ref{eq:Th}), we explicitly write ${\Psi _u}$ as a function of variables $\beta_r$, $\beta_t$, ${\alpha _r}$, and ${\alpha _t}$, which are to be optimized. }
\begin{small}
\begin{align}
&{{\bf P}_1}:\left\{ {\beta _r^*,\beta _t^*,\alpha _r^*,\alpha _t^*} \right\} \!=\! \arg \! \mathop {\max }\limits_{{\beta _r},{\beta _t},{\alpha _r},{\alpha _t}}\!\! {\Psi _u}\left( {{\beta _r},{\beta _t},{\alpha _r},{\alpha _t}} \right) \label{YY}\\
&s.t.\ \ C1:0 < {\beta _r},{\beta _t} < 1, \tag{\ref{YY}{a}} \label{YYa}\\
&\ \ \ \ \ \ C2:{\beta _r} + {\beta _t} = 1,\tag{\ref{YY}{b}} \label{YYb}\\
&\ \ \ \ \ \ C3:P_{{\rm{out}}}^{r,u} < P_{{\rm{out}}}^{r,th},\tag{\ref{YY}{c}} \label{YYc}\\
&\ \ \ \ \ \ C4:P_{{\rm{out}}}^{t,u} < P_{{\rm{out}}}^{t,th},\tag{\ref{YY}{d}} \label{YYd}\\
&\ \ \ \ \ \ C5:JFI > JF{I^{th}},\tag{\ref{YY}{e}} \label{YYe}\\
&\ \ \ \ \ \ C6:0 < {\alpha _r},{\alpha _t} < 1,\tag{\ref{YY}{f}} \label{YYf}\\
&\ \ \ \ \ \ C7:{\alpha _r} + {\alpha _t} = 1.\tag{\ref{YY}{g}} \label{YYg}
\vspace{-0.55cm}
\end{align}
\end{small}%
Constraint (\ref{YYa}) ensures that the amplitude coefficients of transmission and reflection
for the STAR-RIS are between $0$ and $1$.
Constraint (\ref{YYb}) is given to meet the law of energy conservation.
Constraint (\ref{YYc}) and constraint (\ref{YYd}) guarantee that the uplink joint outage probabilities at ${U_r}$ and ${U_t}$ are less than the present thresholds $P_{\text{out}}^{r,th}$ and $P_{\text{out}}^{t,th}$, respectively.
Constraint (\ref{YYe}) ensures that JFI is more than the preset threshold $JF{I^{th}}$.
Constraint (\ref{YYf}) and constraint (\ref{YYg}) specify the range of ${\alpha _r}$, ${\alpha _t}$ and their summation, respectively.

\vspace{-0.45cm}
\subsection{Proposed PSO-PA Algorithm}
Since the sum throughput expression in (\ref{eq:Th}) is intractable, we consider a derivative-free optimization method
to obtain the optimal solution, which is practical without the computation of gradients\cite{9691347}. It is difficult to solve the optimization problem using the exhaustive search method in an acceptable time with the increase in network scale.
In this context, there are low-complexity optimization algorithms, such as genetic algorithm (GA) and particle swarm optimization (PSO), which have been shown to reach the global optimal solution\cite{9785881}.
Although the GA could work as effectively as PSO, PSO offers a superior
computational cost and speed of convergence\cite{Shabir2016ACS,9785881}. Hence, PSO is considered to optimize the energy-splitting ratios ${\beta _r}$, ${\beta _t}$ and  the transmit power
allocation factors ${\alpha _r}$, ${\alpha _t}$ for maximizing the sum throughput.

The proposed PSO-PA method first initializes the particles' positions and velocities. Here, there are four parameters for optimization. The particles' position and velocity can be denoted as ${\bf{d}}_b^w = \left[ {d_{b,1}^w,d_{b,2}^w,d_{b,3}^w,d_{b,4}^w} \right]$ and ${\bf{v}}_b^w = \left[ {v_{b,1}^w,v_{b,2}^w,v_{b,3}^w,v_{b,4}^w} \right]$, where $b$ denotes the particle index and $w$ denotes the iteration index.
Let ${N_p}$ denote the number of particles, and ${N_i}$ denote the number of iterations.
One has ${{\cal D}^w} = \left\{ {{\bf{d}}_1^w, \ldots ,{\bf{d}}_{{N_p}}^w} \right\}$.
Each particle updates the velocity $v_{b,\zeta  }^{w + 1}$ based on its current position $d_{b,\zeta}^{w}$, best record ${{\hat p}_{b,\zeta }^w}$ and global
best record ${{\hat g}_\zeta ^w}$, where $\zeta  \in \left\{ {1,2,3,4} \right\}$. Similarly, one has ${{\hat {\cal P}}^w} = \left\{ {{\bf{\hat p}}_1^w, \ldots ,{\bf{\hat p}}_{{N_p}}^w} \right\}$ and $\hat {\cal G} = \left\{ {{{{\bf{\hat g}}}^1}, \ldots ,{{{\bf{\hat g}}}^{{N_i}}}} \right\}$.
The iterative equation is expressed as
\vspace{-0.15cm}
\begin{small}
\begin{align}
\label{eq:V}
v_{b,\zeta }^{w + 1} \!=\! \bar \omega v_{b,\zeta }^w \!+\! {C_1}{r_1}\left( {\hat p_{b,\zeta }^w - d_{b,\zeta }^w} \right) \!+\! {C_2}{r_2}\left( {\hat g_\zeta ^w - d_{b,\zeta }^w} \right),
\vspace{-0.20cm}
\end{align}
\end{small}%
where ${\bar \omega }$ is an inertia weight,
${r_1}$ and ${r_2}$ are chosen independently and uniformly at random to increase the search randomness, and ${C_1}$ and ${C_2}$ are acceleration instants to control the direction of convergence of the PSO.
For example, a larger value of ${C_1}$ tends to converge to a personal best solution, leading to a local optimal solution. Correspondingly, a larger value of ${C_2}$ tends to converge to the global optimum, where the particles tend to move further distances and increase their exploration of search space.
Each particle updates the position based on its current position $d_{b,\zeta }^{w}$, and the velocity $v_{b,\zeta }^{w + 1}$, as follows
\vspace{-0.25cm}
\begin{small}
\begin{align}
\label{eq:D}
d_{b,\zeta }^{w + 1} = d_{b,\zeta }^w + v_{b,\zeta }^{w + 1}.
\vspace{-0.3cm}
\end{align}
\end{small}%
Each particle evaluates its penalty fitness function based on its positions at
each iteration. The penalty fitness function is given as follows
\vspace{-0.25cm}
\begin{small}
\begin{align}
{P_f} &= {\Psi _u} \!-\! {\mathchar'26\mkern-10mu\lambda _1}\max \left(\! {0,P_{{\rm{out}}}^{r,u} \!-\! P_{{\rm{out}}}^{r,th}}\! \right) \!-\! {\mathchar'26\mkern-10mu\lambda _1}\max \left( \! {0,P_{{\rm{out}}}^{t,u} \!-\! P_{{\rm{out}}}^{t,th}} \!\right)\nonumber\\
& + {\mathchar'26\mkern-10mu\lambda _2}\min \left( {0,JFI - JF{I^{th}}} \right),
\end{align}
\end{small}%
where ${\mathchar'26\mkern-10mu\lambda _1}$ and ${\mathchar'26\mkern-10mu\lambda _2}$ denote the penalty factors for joint outage probability and JFI, respectively.
The proposed PSO-PA scheme is presented in Algorithm 1.

\begin{algorithm}
\begin{small}
\begin{spacing}{1}
    \caption{Particle Swarm Optimization Based Power Allocation (PSO-PA)}
    \label{alg:AOA}
    \renewcommand{\algorithmicrequire}{\textbf{Input:}}
    \renewcommand{\algorithmicensure}{\textbf{Output:}}
    \begin{algorithmic}[1]
        \REQUIRE  ${N_p}$, ${N_i}$, ${C_1}$, and ${C_2}$. 
        \ENSURE The optimal ${\beta _r^*,\beta _t^*,\alpha _r^*,\alpha _t^*}$.
        \STATE Randomly initialize the particle positions $\mathbf{d}_b^1$ and velocities $\mathbf{v}_b^1$, where $b = 1,...,{N_p}$.
        \STATE Set ${\mathbf{\hat {\cal P}^1}} = {\cal D}^1$ and $\mathbf{{\hat g^1}} = \arg \mathop {\max }\limits_{{\mathbf{{\hat p}}_b^1} \in {{{\bf{\hat {\cal P}}}}^{\bf{1}}}}{P_f}$.
        \FOR{ $w = 1:{N_i}$}
        \FOR{$b = 1:{N_p}$}
        \STATE Calculate particle velocity $\mathbf{v}_b^{w + 1}$ via (\ref{eq:V}).
        \STATE Update particle position $\mathbf{d}_b^{w + 1}$ via (\ref{eq:D}).
        \STATE \textbf{if} {${P_f}\left( {\mathbf{d}_b^{w + 1}} \right) > {P_f}\left( {\mathbf{{\hat p}}_b^w} \right)$}
         \STATE \textbf {then} $\mathbf{{\hat p}}_b^{w + 1} = \mathbf{d}_b^{w + 1}$.
         \STATE \textbf{else}  $\mathbf{{\hat p}}_b^{w + 1} = \mathbf{{\hat p}}_b^w$.
         \STATE \textbf{end if}
         \ENDFOR
         \STATE Set $\mathbf{{\hat g}}^{w + 1} = \arg \mathop {\max }\limits_{\mathbf{{\hat p}}_b^{w + 1} \in {\bf{{\hat {\cal P}}}}^{w + 1}} {P_f}$.
         \ENDFOR
         \STATE ${\left( {\mathbf{{\hat g}}^w} \right)^*} = \arg \mathop {\max }\limits_{\mathbf{{\hat g}}^w \in {\bf{{\hat {\cal G}}}}}{P_f}$.
    \end{algorithmic}
    \end{spacing}
    \end{small}
\end{algorithm}

The time complexity of PSO-PA can be calculated as follows. In PSO-PA, given the ${N_p}$ and ${N_i}$,
the time complexity for initializing the particles' positions and velocities can be expressed as $O\left( {N_p} \right)$. The time complexity for searching for the global best record is given by $O\left( {{N_i} \times {N_p}} \right)$. Hence, the overall time complexity is approximated as $O\left( {{N_i} \times {N_p}} \right) + O\left( {N_p} \right) \approx O\left( {{N_i} \times {N_p}} \right)$.

\vspace{-0.35cm}
\section{Results and Discussion} 	

\begin{table}[t]
 \centering
  \caption{SYSTEM PARAMETERS}
  \label{table:Table I}
  \begin{tabular}{  m{4.7cm}<{\centering} | m{2.9cm}<{\centering}  }
  \Xhline{1pt}
    The transmit power at the AP  &  ${P_{{\rm{AP}}}} = 1$ Watt (W) \\
  \Xhline{0.5pt}
  The Nakagami-$m$ fading channel parameters & ${m_{0,k}} = {m_{i,\ell }} = {m_{k,i}} = 2$  \\
  \Xhline{0.5pt}
  The symbol duration&  ${T_s} = \left( {1/256000} \right){\rm{s}}$ \cite{9464720}  \\
    \Xhline{0.5pt}
    The energy conversion  efficiency factor & $\eta  = 0.9$   \\
  \Xhline{0.5pt}
    The initial state of the battery &  $C = 1~\rm{\mu J}$  \\
   \Xhline{0.5pt}
   The number of time slots & $Q=5$   \\
  \Xhline{0.5pt}
    The path-loss at a reference  distance of $1$ meter &  $\tau =  - 2$~dB   \\
  \Xhline{0.5pt}
    The path-loss exponent & $\vartheta  = 2$   \\
  \Xhline{0.5pt}
      The number of antennas & $L = 2$   \\
  \Xhline{1pt}
  \end{tabular}
  \vspace{-0.5cm}
\end{table}

In this section, power outage probability, information outage probability, sum throughput, and joint outage probability of the proposed system over Nakagami-$m$ fading channels are evaluated.
The main simulation parameters are listed in Table~I; the other parameters are specified in figure caption.
We consider a two-dimensional Cartesian coordinate location setup as follow. The AP is located at the origin $\left( {0,0} \right)$, and the STAR-RIS is located at $\left( {16,0} \right)$. The location of $U_t$ is $\left( {16,2} \right)$, and that of $U_r$ is $\left( {16, - 3} \right)$.
We assume that the inter-distance between the adjacent elements of STAR-RIS is half-wavelength. Accordingly, reflecting/transmitting channels through the adjacent elements are regarded as independent channels \cite{9834288}.
To verify the effectiveness of the proposed system, the following benchmarks are considered for comparison: i) conventional RIS case. More specifically, we deploy the transmitting-only RIS and the reflecting-only RIS at the same location as the STAR-RIS for full-space coverage, where each reflecting/transmitting-only RIS is equipped with $\frac{N}{2}$ elements for fairness comparison\cite{9834288}; ii) TDMA case. More specifically, the STAR-RIS adopts
the time-switching protocol; iii) buffer-less case; iv) discrete phase-shift case. Because of hardware limitations, discrete phase-shift adjustment is practical and realistic for STAR-RIS\cite{10040906, 10439654}. Specifically, discrete phase-shift values are obtained by uniformly quantizing the feasible region \cite{10040906}, i.e.,  $\varphi _i^r,\varphi _i^t \in \Im  = \left\{ {0,\frac{{2\pi }}{{{2^e}}}, \ldots ,\frac{{2\pi \left( {{2^e} - 1} \right)}}{{{2^e}}}} \right\}$, where $e$ denotes the number of quantization bits and is set to $4$ in this paper. Hence, baseline schemes include the proposed system with discrete phase shifts (referred to as discrete phase),
the STAR-RIS aided MISO SWIPT-NOMA buffer-less system (referred to as STAR-RIS-BL-NOMA), the RIS and energy buffer aided MISO SWIPT-NOMA system (referred to as RIS-EB-NOMA), and the STAR-RIS and energy buffer aided MISO SWIPT-TDMA system (referred to as STAR-RIS-EB-TDMA).
It is noted that the proposed system and STAR-RIS-BL-NOMA have the same information transmission scheme, i.e., STAR-RIS aided NOMA. Hence, we only consider the proposed system, RIS-EB-NOMA, and the STAR-RIS-EB-TDMA when comparing the performance of the information outage probability and sum throughput.
\begin{figure}
\center
\includegraphics[width=3.2in,height=2.4in]{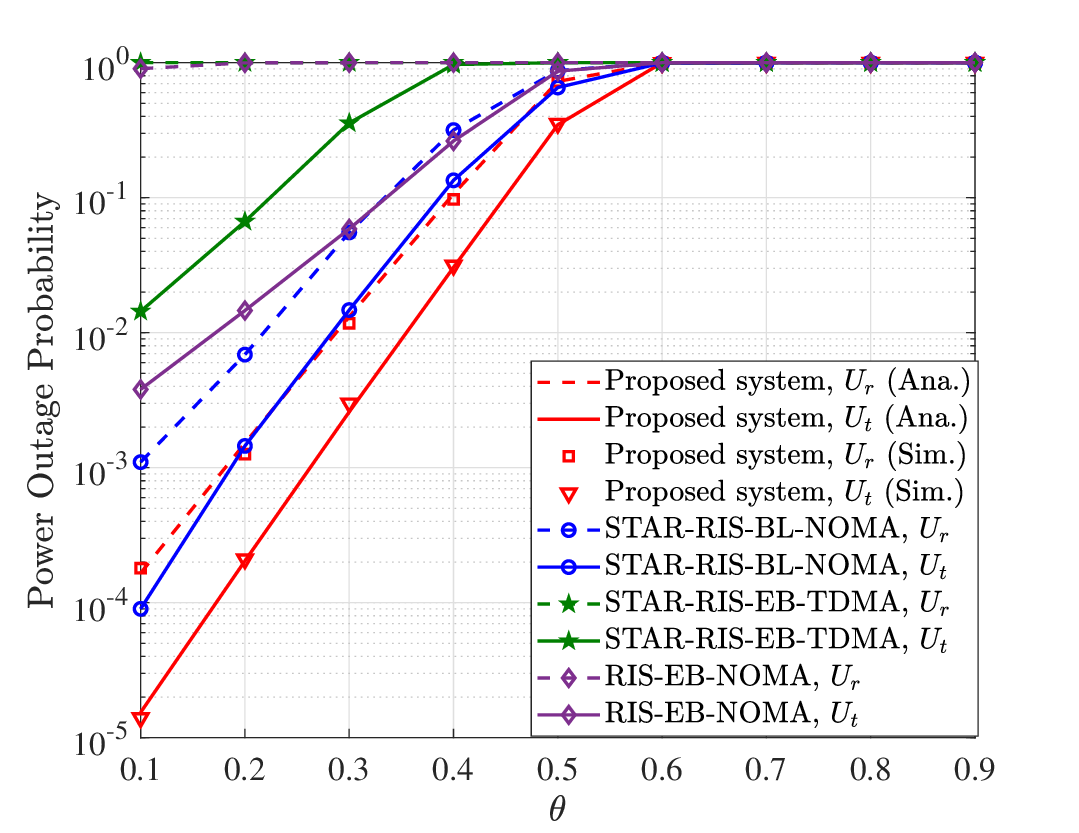}
\vspace{-0.3cm}
\caption{Power outage probability versus power splitting factor $\theta $ of various systems, where $N = 18$, ${\beta _r} = 0.65$, and ${\beta _t} = 0.35$.}
\label{fig:PoutVS}
\vspace{-0.4cm}
\end{figure}
\begin{figure}
\center
\includegraphics[width=3.2in,height=2.4in]{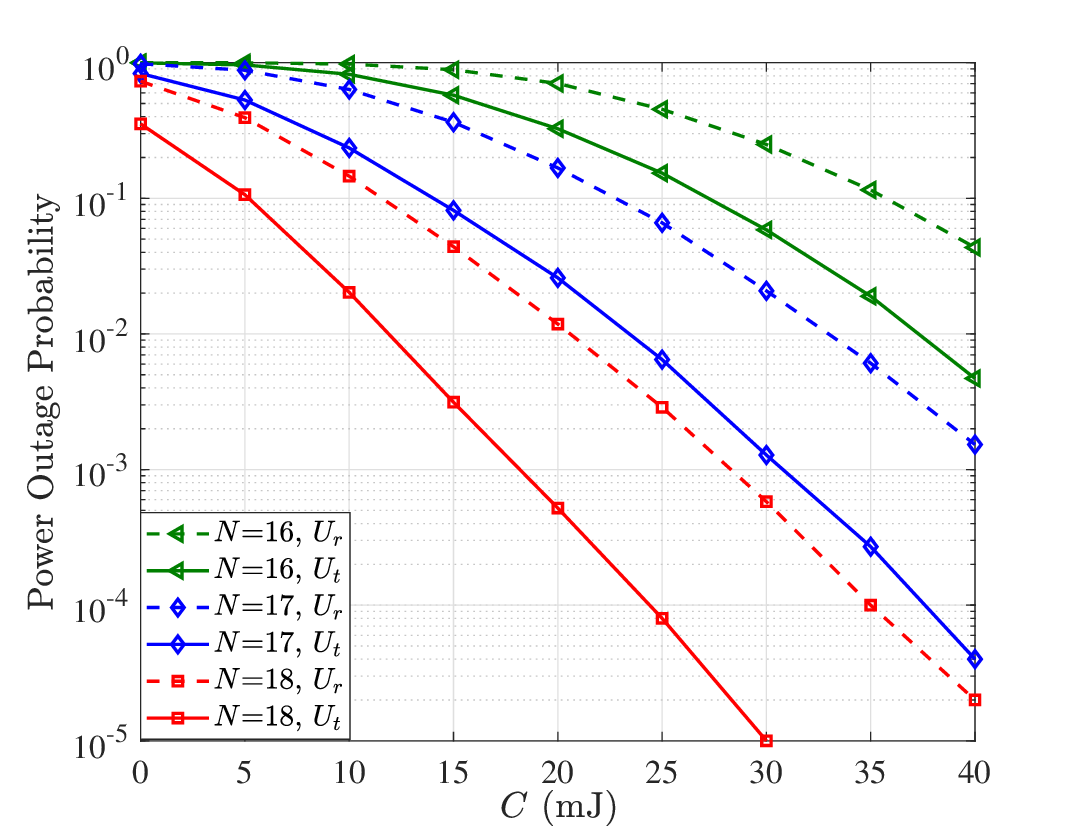}
\vspace{-0.3cm}
\caption{Power outage probability versus the initial energy $C$ for the proposed system with various values of $N$, where ${\beta _r} = 0.65$, ${\beta _t} = 0.35$, and $\theta  = 0.5$.}
\label{fig:PoutVSbeta}
\vspace{-0.65cm}
\end{figure}

Fig.~\ref{fig:PoutVS} shows the power outage probability versus power-splitting factor $\theta $ of various systems, where $N = 18$, ${\beta _r} = 0.65$, and ${\beta _t} = 0.35$.
It can be observed that the simulated curves well match with the theoretical ones, which verifies the correctness of the proposed analytical method.
Moreover, it is observed that larger values of $\theta $ lead to higher power outage probabilities due to less power being allocated to energy collection.
Furthermore, it is seen that the proposed system provides improved power outage probability performance compared to the STAR-RIS-BL-NOMA, RIS-EB-NOMA, and STAR-RIS-EB-TDMA. For example, when $\theta  = 0.3$, the power outage probability of the proposed system at ${U_r}$ is approximately  $80\%$ lower than that of the STAR-RIS-BL-NOMA.
It is also seen that the proposed system yields an improved performance compared to the RIS-EB-NOMA.
This is because the STAR-RIS can be configured for full-space electromagnetic propagation environments, whereas the conventional RIS requires both the transmitting-only RIS and the reflecting-only RIS to provide full-space coverage.
It is shown that the proposed system outperforms the STAR-RIS-EB-TDMA. The reason is that ${U_r}$ and ${U_t}$ can utilize the same block of time-frequency resources in NOMA, thus achieving a multiplexing gain compared with the TDMA.

Fig.~\ref{fig:PoutVSbeta} illustrates the power outage probability versus the initial energy $C$ for the proposed system with various values of $N$, where ${\beta _r} = 0.65$, ${\beta _t} = 0.35$, and $\theta  = 0.5$.
It is observed that the increase in the number of STAR-RIS elements significantly improves the power outage probability performance{\footnote{When the number of reflective elements becomes large, the circuit power consumption of STAR-RIS grows accordingly\cite{9615187}. The study of minimizing reflective elements while meeting system performance constraints will be left for future work.}}.
For example, when $C = 15$~mJ, the outage probability at $U_r$ for $N=18$ is approximately $85\%$ lower than
that of $N=17$ while the outage probability at $U_r$ for $N=17$ is approximately $60\%$ lower than that case where $N=16$.
It is also observed that an increase in $C$ yields a better power outage probability performance, especially when $N$ is larger. For example, when $N=18$, an increase in $C$ from $20$ to $25$ reduces the outage probability of $U_t$ from ${10^{ - 3}}$ to ${10^{ - 4}}$.
\begin{figure}
\center
\includegraphics[width=3.2in,height=2.4in]{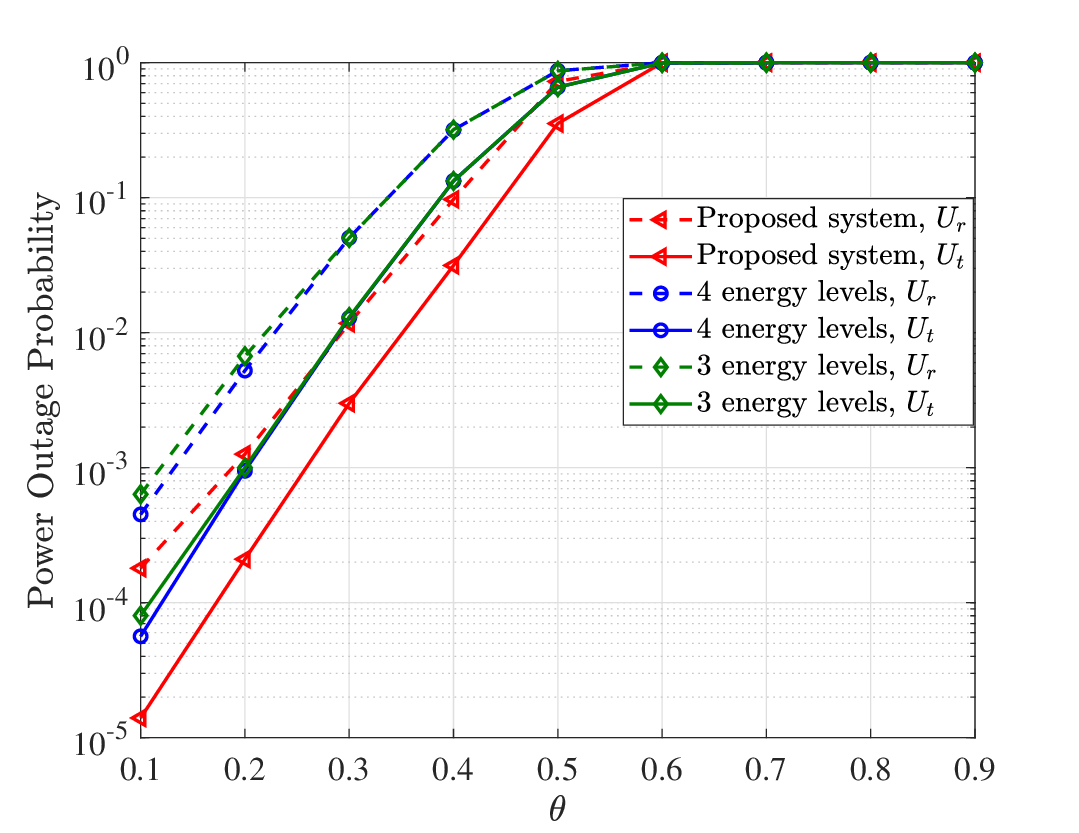}
\vspace{-0.3cm}
\caption{Power outage probability versus power splitting factor $\theta $ of different energy
levels, where $N = 18$, ${\beta _r} = 0.65$, and ${\beta _t} = 0.35$.}
\label{fig:buffer_state}
\vspace{-0.5cm}
\end{figure}
\begin{figure}
\center
\includegraphics[width=3.2in,height=2.4in]{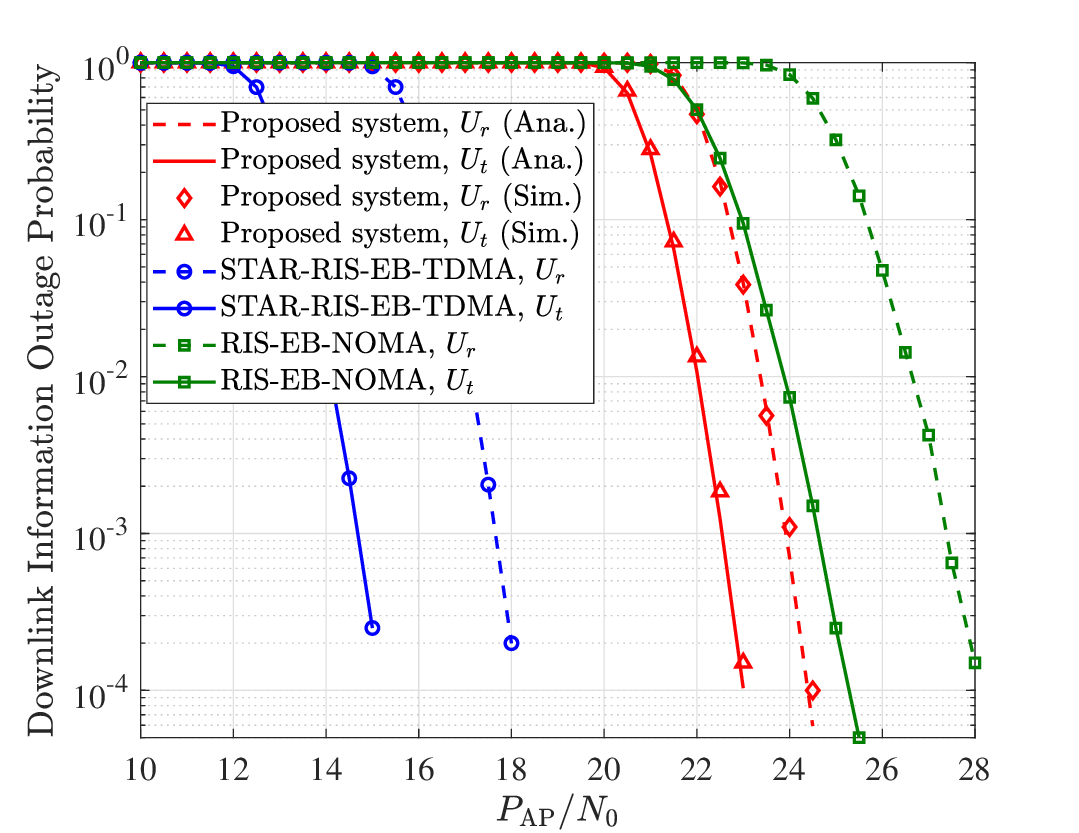}
\vspace{-0.3cm}
\caption{Downlink information outage probability versus the transmit SNR of various systems, where $N = 18$, $\theta  = 0.5$, ${\beta _r} = 0.6$, ${\beta _t} = 0.4$, ${\alpha _r} = 0.6$, ${\alpha _t} = 0.4$, and $R = 1$.}
\label{fig:IVS}
\vspace{-0.4cm}
\end{figure}

Fig.~\ref{fig:buffer_state} plots the power outage probability versus power splitting factor $\theta $ of different energy levels, where $N = 18$, ${\beta _r} = 0.65$, and ${\beta _t} = 0.35$. For the benchmark schemes, the energy buffer is discretized into $3$ and $4$ possible energy levels, respectively. It is observed that the benchmark schemes perform worse than the proposed system. The reason behind this is that the energy level of the energy buffer is rounded down to the closest discrete level\cite{8105867}, which causes a partial loss of energy. Moreover, it is seen that the performance of $4$ energy levels is better than that of $3$ energy levels due to the larger buffer capacity.
Hence, unlike\cite{9374454} which performs a residual energy-aware model to extend the lifetime of the node and\cite{8105867} which schedules nodes to reduce packet loss based on their energy levels and data queue lengths, a two-state Markov chain is suitable for modeling power outage and achieving better performance.

Fig.~\ref{fig:IVS} depicts the downlink information outage probability versus the transmit SNR of various systems, where $N = 18$, $\theta  = 0.5$, ${\beta _r} = 0.6$, ${\beta _t} = 0.4$, ${\alpha _r} = 0.6$, ${\alpha _t} = 0.4$, and $R = 1$.
It can be seen that the STAR-RIS-EB-TDMA outperforms the proposed system in information outage probability. The reason is that there is no inter-user interference with STAR-RIS-EB-TDMA.
Moreover, it can be seen that the proposed system outperforms the RIS-EB-NOMA. For example, for a target information outage probability of ${10^{ - 3}}$, the proposed system shows a gain of $3.5$~dB
compared to RIS-EB-NOMA for $U_r$.

Fig.~\ref{fig:SumthVS} shows the downlink sum throughput versus transmit SNR of various systems, where $N = 18$, $\theta  = 0.5$, ${\beta _r} = 0.65$, ${\beta _t} = 0.35$, ${\alpha _r} = 0.95$, ${\alpha _t} = 0.05$, and $R = 1.5$.
It is observed that the proposed system can provide an improved sum throughput performance compared with baseline schemes.
This is because compared to TDMA, the ${U_r}$ and ${U_t}$ can utilize the same block of time-frequency resources
in NOMA to obtain a multiplexing gain.
Moreover, in NOMA, the STAR-RIS adopts the energy-splitting protocol and exploit the proper energy-splitting ratios to obtain an enhanced performance.
At low SNR, the performance of the proposed system is slightly worse than that of STAR-RIS-EB-TDMA. The reason behind this is that the poor information outage performance of the proposed system at low SNR affects the sum throughput performance.
Moreover, it is observed that an intersection exists between the curves of RIS-EB-NOMA and STAR-RIS-EB-TDMA.
The reason lies that the beamforming gain of the STAR-RIS takes the lead and the multiplexing gain is not obvious at low SNR, whereas the multiplexing gain enhances and ultimately surpasses the beamforming gain as a leading factor at high SNR.

\begin{figure}
\center
\includegraphics[width=3.2in,height=2.4in]{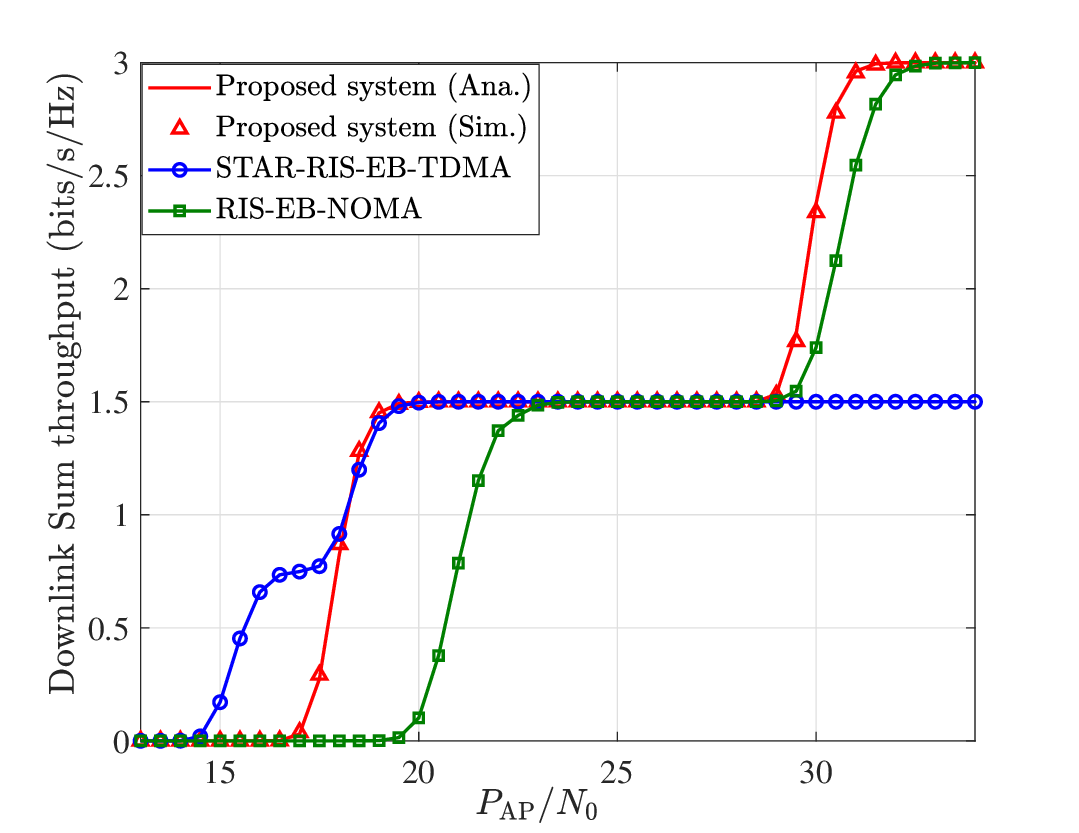}
\vspace{-0.3cm}
\caption{Downlink sum throughput versus transmit SNR of various systems, where $N = 18$, $\theta  = 0.5$, ${\beta _r} = 0.65$, ${\beta _t} = 0.35$, ${\alpha _r} = 0.95$, ${\alpha _t} = 0.05$, and $R = 1.5$.}
\label{fig:SumthVS}
\vspace{-0.36cm}
\end{figure}
\begin{figure}
\center
\includegraphics[width=3.2in,height=2.4in]{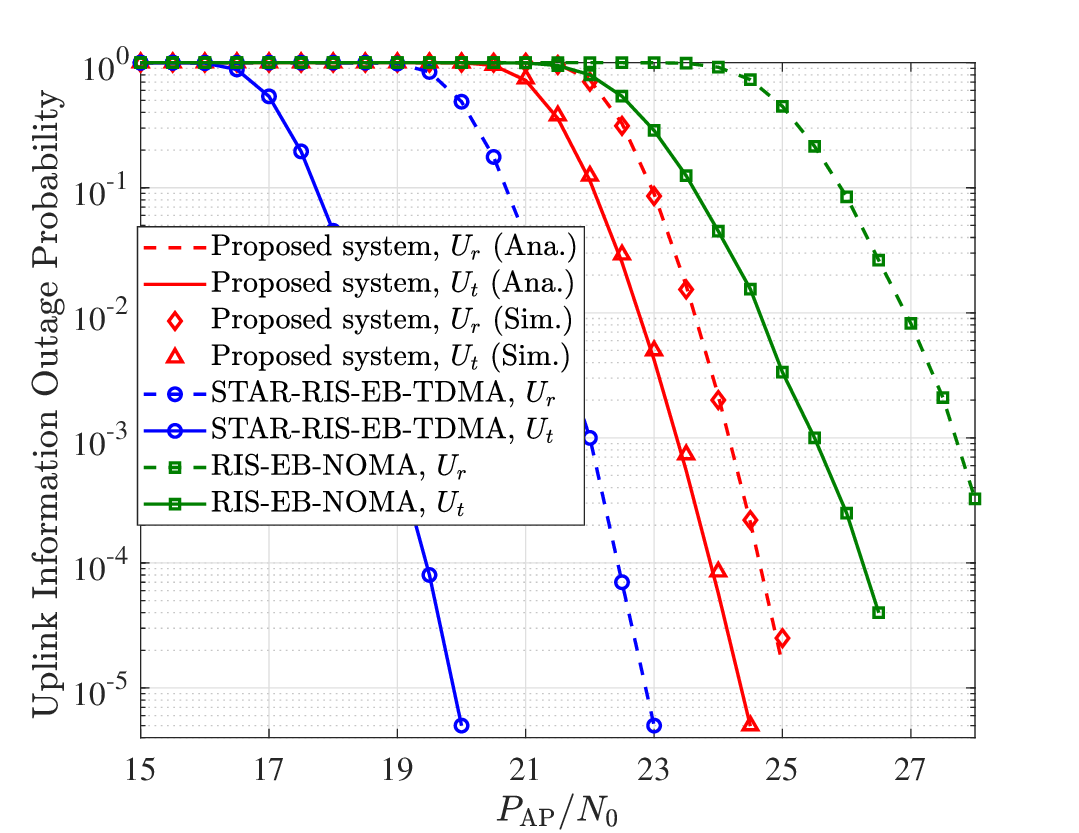}
\vspace{-0.3cm}
\caption{Uplink information outage probability versus the transmit SNR of various systems, where $N = 18$, ${\beta _r} = 0.6$, ${\beta _t} = 0.4$, and $R = 0.1$.}
\label{fig:I_up_pout}
\vspace{-0.45cm}
\end{figure}
\begin{figure}
\center
\includegraphics[width=3.2in,height=2.4in]{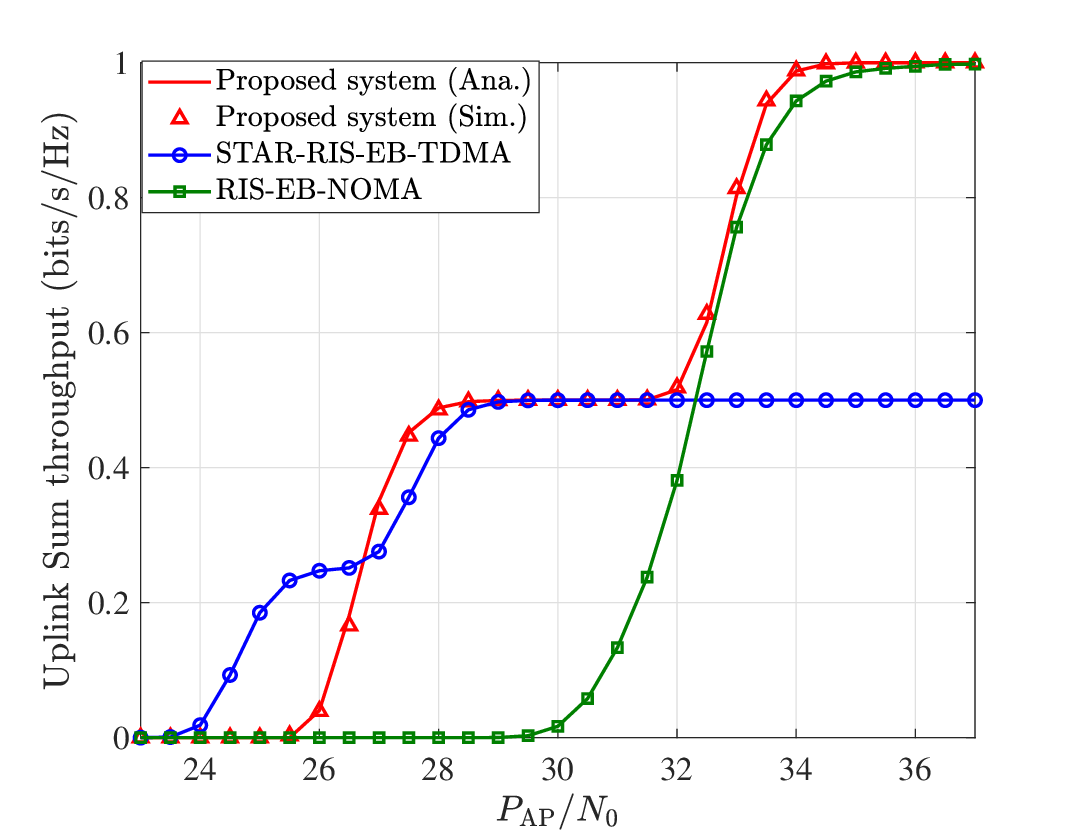}
\vspace{-0.3cm}
\caption{Uplink sum throughput versus transmit SNR of various systems, where $N = 18$, ${\beta _r} = 0.3$, ${\beta _t} = 0.7$, and $R = 0.5$.}
\label{fig:I_up_th}
\vspace{-0.45cm}
\end{figure}
\begin{figure}
\center
\includegraphics[width=3.2in,height=2.4in]{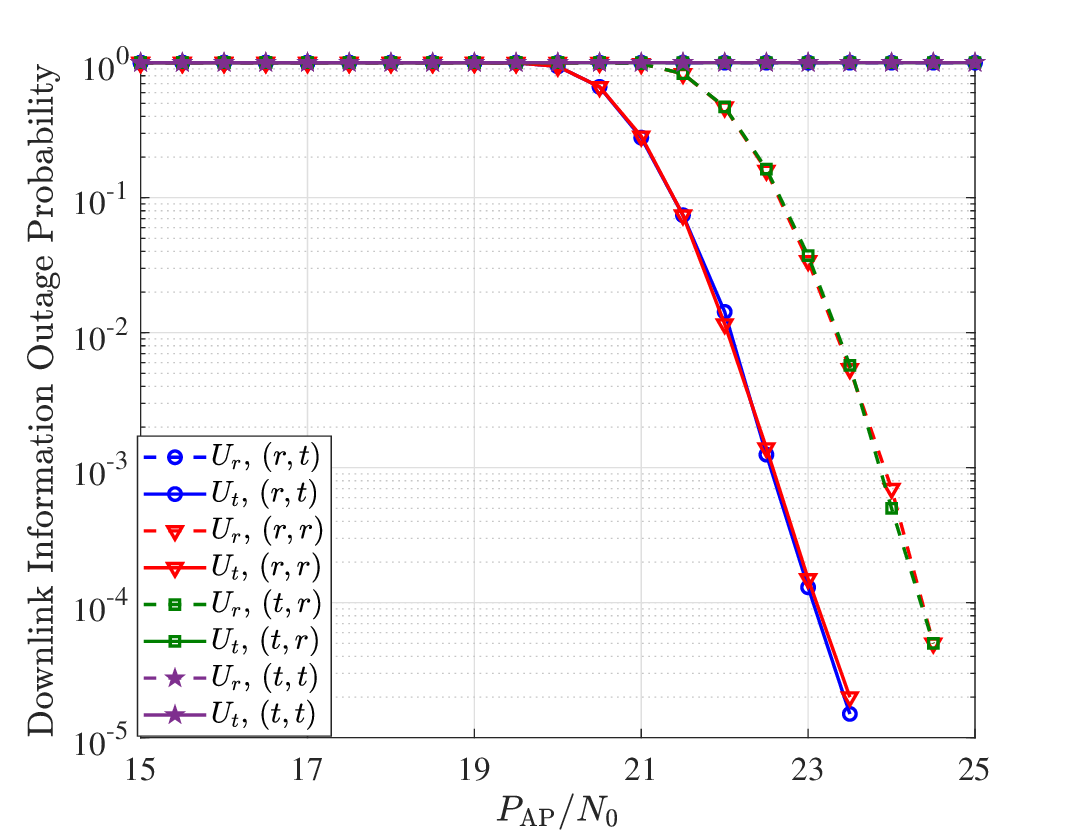}
\vspace{-0.3cm}
\caption{Downlink information outage probability versus the transmit SNR of various decoding orders, where $N = 18$, $\theta  = 0.5$, ${\beta _r} = 0.6$, ${\beta _t} = 0.4$, ${\alpha _r} = 0.6$, ${\alpha _t} = 0.4$, and $R = 1$.}
\label{fig:decoding_order}
\vspace{-0.4cm}
\end{figure}

Fig.~\ref{fig:I_up_pout} illustrates the uplink information outage probability versus the transmit SNR of various systems, where $N = 18$, ${\beta _r} = 0.6$, ${\beta _t} = 0.4$, and $R = 0.1$. Similar to the downlink information outage probability, the uplink information outage performance of the STAR-RIS-EB-TDMA is better than that of the proposed system, while the proposed system outperforms the RIS-EB-NOMA.
Fig.~\ref{fig:I_up_th} depicts the uplink sum throughput versus transmit SNR of various systems, where $N = 18$, ${\beta _r} = 0.3$, ${\beta _t} = 0.7$, and $R = 0.5$. It can be observed that NOMA schemes have better sum throughput performance compared to the TDMA scheme due to the multiplexing gain, and the proposed system provides superior performance over RIS-EB-NOMA.

Fig.~\ref{fig:decoding_order} shows the downlink information outage probability versus the transmit SNR of various decoding orders, where $N = 18$, $\theta  = 0.5$, ${\beta _r} = 0.6$, ${\beta _t} = 0.4$, ${\alpha _r} = 0.6$, ${\alpha _t} = 0.4$, and $R = 1$.
For the case of two users, four decoding orders are considered according to\cite{9188014}. In the decoding order $(i,j)$ of the legend, $i$ denotes that ${U_i}$'s data will be decoded first by $U_t$, and $j$ denotes that ${U_j}$'s data will be decoded first by $U_r$, where $i,j \in \{ t,r\} $.
For example, the decoding order $\left( {r,r} \right)$ means that ${U_t}$ firstly decodes the signal of ${U_r}$, and ${U_r}$ directly decodes the signal of ${U_r}$. It can be seen that for the decoding order $\left( {r,t} \right)$, ${U_r}$ firstly decodes the signal of ${U_t}$, resulting in constant outage. The reason is that due to the transmit
power allocation ${\alpha _r} > {\alpha _t}$, if ${U_r}$ firstly decodes the signal of ${U_t}$ by treating the signal of ${U_r}$ as interference, the decoding performance of ${U_r}$ will become worse. It is also seen that for the decoding order $\left( {t,r} \right)$, ${U_t}$ cannot decode the signal of ${U_t}$ first by treating the signal of ${U_r}$ as interference. Hence, the decoding order $\left( {r,r} \right)$ is the optimal decoding order.

\begin{figure}
\center
\includegraphics[width=3.2in,height=2.4in]{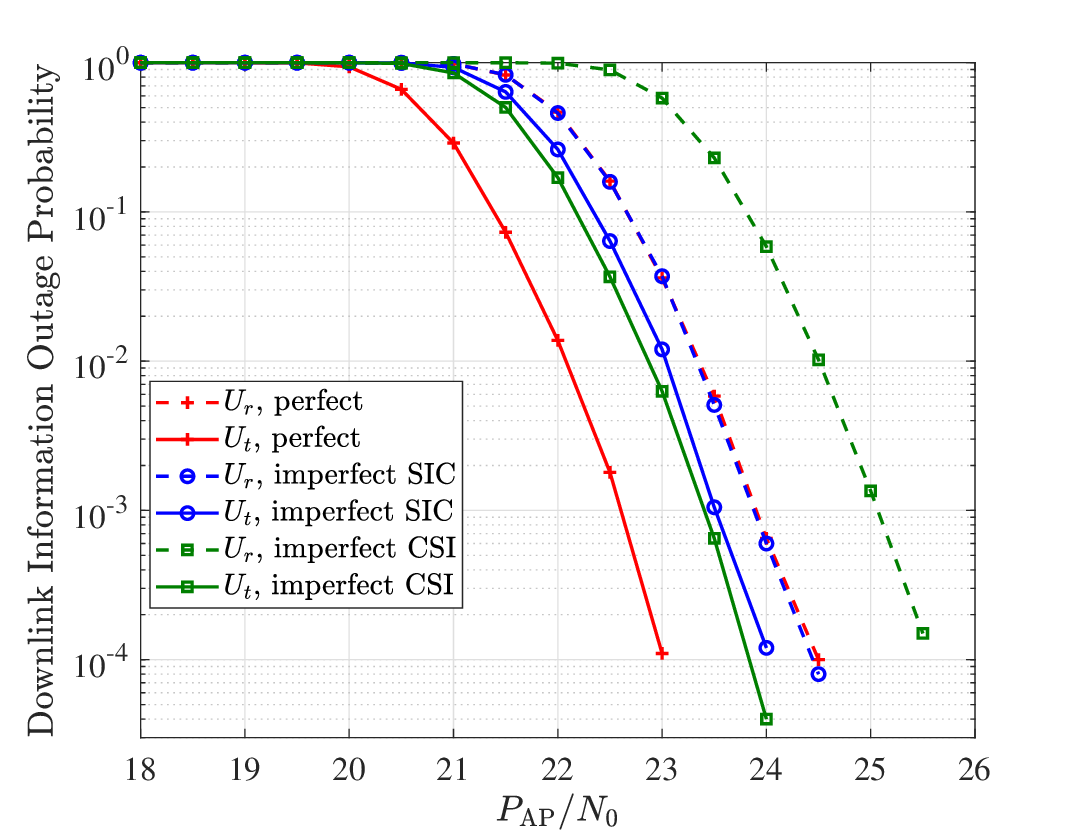}
\vspace{-0.3cm}
\caption{Downlink information outage probability versus the transmit SNR of imperfect SIC and imperfect CSI, where $N = 18$, $\theta  = 0.5$, ${\beta _r} = 0.6$, ${\beta _t} = 0.4$, ${\alpha _r} = 0.6$, ${\alpha _t} = 0.4$, and $R = 1$.}
\label{fig:imperfect}
\vspace{-0.4cm}
\end{figure}
\begin{figure}
\center
\includegraphics[width=3.2in,height=2.4in]{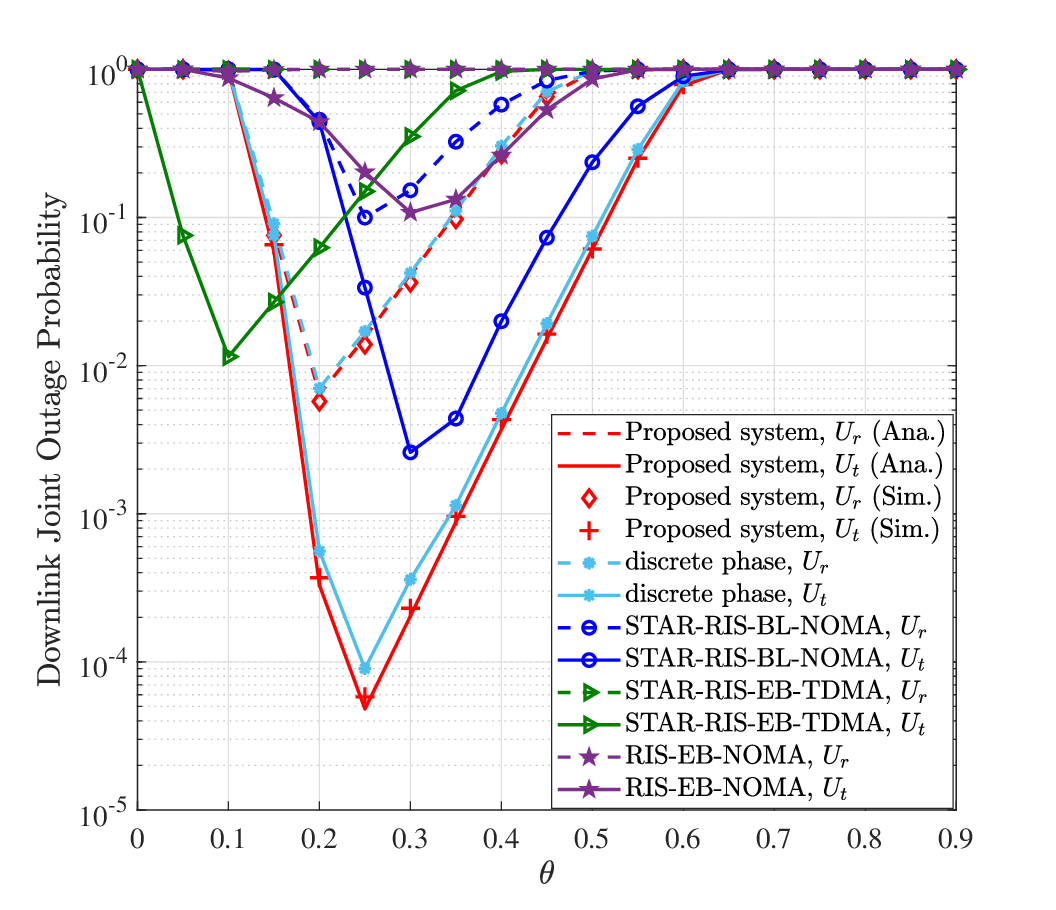}
\vspace{-0.3cm}
\caption{Downlink joint outage probability versus power splitting factor $\theta $ of various systems, where $N = 18$, ${P_{{\rm{AP}}}}/{N_0} = 25$~dB, ${\beta _r} = 0.6$, ${\beta _t} = 0.4$, ${\alpha _r} = 0.7$, and ${\alpha _t} = 0.3$.}
\label{fig:thVSR}
\vspace{-0.45cm}
\end{figure}

Fig.~\ref{fig:imperfect} depicts the downlink information outage probability versus the transmit SNR of imperfect SIC and imperfect CSI, where $N = 18$, $\theta  = 0.5$, ${\beta _r} = 0.6$, ${\beta _t} = 0.4$, ${\alpha _r} = 0.6$, ${\alpha _t} = 0.4$, and $R = 1$. For imperfect SIC, there will be residual interference after performing the SIC. Thus, ${U_t}$ cannot completely remove the interference from the signal of ${U_r}$. The SNR in (\ref{eq:rdt}) at $U_t$ can be re-expressed as
\vspace{-0.25cm}
\begin{small}
\begin{align}
\gamma _t^d = \frac{{\theta {\beta _t}{{\left| {\sum\limits_{\ell  = 1}^L {\left( {h_{t,\ell }^0 + \sum\limits_{i = 1}^N {{h_{\ell ,i}}{g_{t,i}}} } \right)} } \right|}^2}{\alpha _t}{P_{{\rm{AP}}}}}}{{\varepsilon \theta {\beta _t}{{\left| {\sum\limits_{\ell  = 1}^L {\left( {h_{t,\ell }^0 + \sum\limits_{i = 1}^N {{h_{\ell ,i}}{g_{t,i}}} } \right)} } \right|}^2}{\alpha _r}{P_{{\rm{AP}}}} + L{N_0}}},
\vspace{-0.25cm}
\end{align}
\end{small}%
where $\varepsilon  \in \left[ {0,1} \right]$ denotes the portion of residual interference. It is seen that the performance of ${U_t}$ decreases due to the residual interference, and that of ${U_r}$ is unchanged since no SIC is performed. This result is consistent with that from Fig.~\ref{fig:imperfect}. For imperfect CSI, the complex channel coefficients are represented as $\hat h_{k,\ell }^0 = \tilde h_{k,\ell }^0 + \bar h_{k,\ell }^0$, ${{\hat h}_{\ell ,i}} = {{\tilde h}_{\ell ,i}} + {{\bar h}_{\ell ,i}}$, and ${{\hat g}_{k,i}} = {{\tilde g}_{k,i}} + {{\bar g}_{k,i}}$, where $\tilde h_{k,\ell }^0$, ${{\tilde h}_{\ell ,i}}$, ${{\tilde g}_{k,i}}$ are the estimated channel coefficients, and $\bar h_{k,\ell }^0$, ${{\bar h}_{\ell ,i}}$, ${{\bar g}_{k,i}}$ are the estimation errors\cite{9854887,10129204}. Hence, (\ref{eq:yk}) can be re-expressed as
\vspace{-0.25cm}
\begin{small}
\begin{align}
{y_k} &\!=\! \sqrt {\theta {\beta _k}} \sum\limits_{\ell  = 1}^L \!\!{\left(\!\!\! \begin{array}{l}
\left( {\tilde h_{k,\ell }^0 + \bar h_{k,\ell }^0} \right){e^{ - j{\varpi _\ell }}}\! + \!\sum\limits_{i = 1}^N \!\!{\left( {{{\tilde h}_{\ell ,i}} + {{\bar h}_{\ell ,i}}} \right)} \nonumber\\
 \times \left( {{{\tilde g}_{k,i}} + {{\bar g}_{k,i}}} \right){e^{j\left( {\varphi _i^k - {\varpi _\ell }} \right)}}
\!\!\end{array} \!\!\!\right)}\! s + {n_k}\\
 &= \sqrt {\theta {\beta _k}} \sum\limits_{\ell  = 1}^L {\left( {\tilde h_{k,\ell }^0{e^{ - j{\varpi _\ell }}} + \sum\limits_{i = 1}^N {{{\tilde h}_{\ell ,i}}{{\tilde g}_{k,i}}{e^{j\left( {\varphi _i^k - {\varpi _\ell }} \right)}}} } \right)} s\nonumber\\
 &+ \sqrt {\theta {\beta _k}} {{\bar n}_k} + {n_k},
 \vspace{-0.25cm}
\end{align}
\end{small}%
where
\vspace{-0.25cm}
\begin{small}
\begin{align}
{{\bar n}_k} \!=\! \sum\limits_{\ell  = 1}^L\! {\left(\! {\bar h_{k,\ell }^0{e^{ - j{\varpi _\ell }}} \!+\!\! \sum\limits_{i = 1}^N \!{\left( \!\!\begin{array}{l}
{{\tilde h}_{\ell ,i}}{{\bar g}_{k,i}}\\
 + {{\bar h}_{\ell ,i}}{{\tilde g}_{k,i}} + {{\bar h}_{\ell ,i}}{{\bar g}_{k,i}}
\!\!\end{array} \right)\!\!{e^{j\left( {\varphi _i^k - {\varpi _\ell }}\!\! \right)}}} } \!\!\right)} s
\vspace{-0.25cm}
\end{align}
\end{small}%
is the error due to imperfect CSI. Since ${{\bar n}_k}$ is the sum of a large number of random variables, it can be approximated as a Gaussian-distributed random variable with variance ${{N_0}}$ by using the central limit theorem (CLT)\cite{9854887}. Thus, there will be one interference term ${L\theta {\beta _k}{N_0}}$ in (\ref{eq:rds}), (\ref{eq:rdt}), and (\ref{eq:rdr}). It is observed that different from the case of imperfect SIC, the performance of all users deteriorates due to the imperfect CSI. This result is also consistent with that from Fig.~\ref{fig:imperfect}.

\begin{figure}
\center
\includegraphics[width=3.2in,height=2.4in]{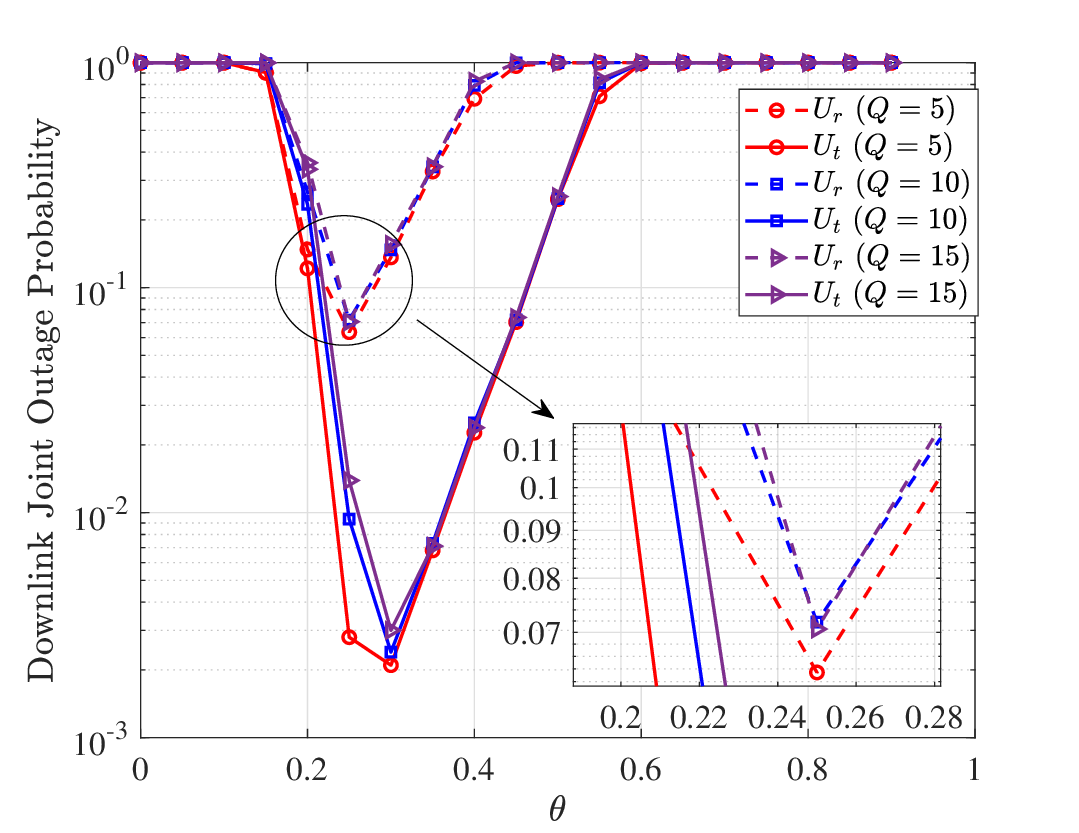}
\vspace{-0.3cm}
\caption{Downlink joint outage probability versus power-splitting factor $\theta $  of the proposed system for various values of $Q$, where $N = 17$, ${P_{{\rm{AP}}}}/{N_0} = 24$~dB, ${\beta _r} = 0.6$, ${\beta _t} = 0.4$, ${\alpha _r} = 0.7$, and ${\alpha _t} = 0.3$.}
\label{fig:thVStime}
\vspace{-0.45cm}
\end{figure}
\begin{figure}
\center
\includegraphics[width=3.2in,height=2.4in]{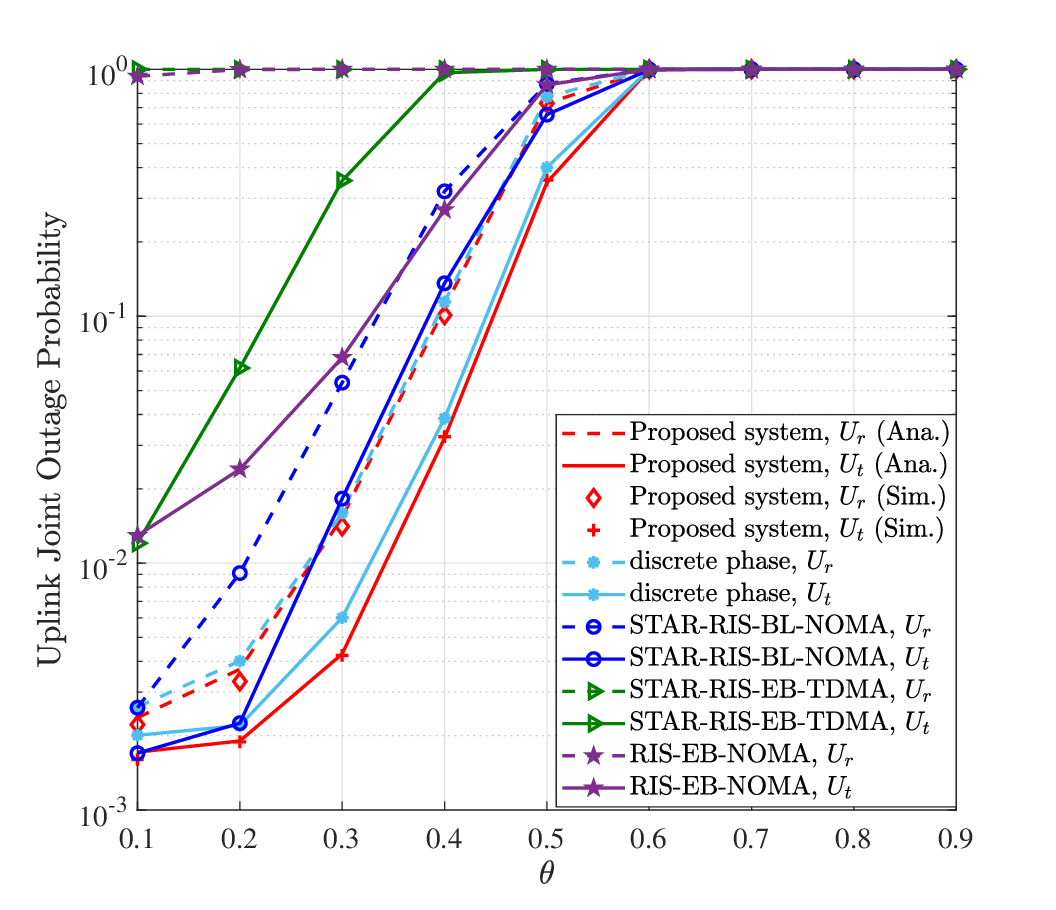}
\vspace{-0.3cm}
\caption{Uplink joint outage probability versus power splitting factor $\theta $ of various systems, where $N = 18$, ${P_{{\rm{AP}}}}/{N_0} = 24$~dB, ${\beta _r} = 0.65$, and ${\beta _t} = 0.35$.}
\label{fig:joint_up_pout}
\vspace{-0.55cm}
\end{figure}

Fig.~\ref{fig:thVSR} plots the downlink joint outage probability versus power splitting factor $\theta $ of various systems, where $N = 18$, ${P_{{\rm{AP}}}}/{N_0} = 25$~dB, ${\beta _r} = 0.6$, ${\beta _t} = 0.4$, ${\alpha _r} = 0.7$, and ${\alpha _t} = 0.3$.
It is noticed that an optimal value of $\theta $ exists, which balances the power and information outage probabilities to minimize the joint outage probability.
It is observed that the proposed system can obtain better joint outage probability performance compared with the baseline schemes.
For example, when $\theta = 0.3$, the joint outage probability of the proposed system at $U_t$ is ${10^{ - 3}}$ while that of the STAR-RIS-BL-NOMA is ${10^{ - 2}}$.
It is also noticed that the optimal value of $\theta $ for $U_t$ is larger than that for $U_r$. This is because $U_t$ is a close user, which experiences better channel conditions than $U_r$ and can allocate more power for information decoding.
Moreover, the STAR-RIS-EB-TDMA slightly outperforms the proposed system for small values of $\theta $.
This is because the STAR-RIS-EB-TDMA has a better information outage probability than the proposed system.
When $\theta $ is small, the poor information outage probability of the proposed system damages the joint outage probability performance. However, with the increase of $\theta $ and the improvement in the information outage probability performance, the proposed system is superior to the STAR-RIS-EB-TDMA scheme.
There is a small performance gap between the proposed system with continuous phase shifts and the proposed system with discrete phase shifts. Hence, 4-bit quantization of discrete phase yields nearly the same performance as continuous phase shifts and phase errors can be neglected \cite{9099280}, which offers more insights into the practical deployment of the proposed system.

Fig.~\ref{fig:thVStime} shows the downlink joint outage probability versus power splitting factor $\theta $  of the proposed system for various values of $Q$, where $N = 17$, ${P_{{\rm{AP}}}}/{N_0} = 24$~dB, ${\beta _r} = 0.6$, ${\beta _t} = 0.4$, ${\alpha _r} = 0.7$, and ${\alpha _t} = 0.3$.
It is observed that the joint outage probability performance decreases with the increase of $Q$. This is because as the number of continuous-time slots $Q$ increases, both information outage and power outage increase.
Moreover, it is observed that as $Q$ increases, the performance degradation becomes smaller.
For example, the gap between the curves of $Q=10$ and $Q=15$ is relatively small.

Fig.~\ref{fig:joint_up_pout} depicts the uplink joint outage probability versus power splitting factor $\theta $ of various systems, where $N = 18$, ${P_{{\rm{AP}}}}/{N_0} = 24$~dB, ${\beta _r} = 0.65$, and ${\beta _t} = 0.35$. It can be observed that the performance of uplink joint outage probability deteriorates as $\theta $ increases, which is different from the downlink joint outage probability. The reason is that larger values of $\theta $ lead to higher power outage probability, while uplink information outage probability is independent of $\theta $. It is seen that the proposed system achieves better performance than STAR-RIS-BL-NOMA.
Since energy buffer is not considered for the STAR-RIS-BL-NOMA, it causes uplink information outage due to lack of energy, thus deteriorating the performance of uplink joint outage probability.
The 4-bit quantization of discrete phase yields nearly the same performance as the proposed system in the uplink joint outage probability.

\begin{figure}
\center
\includegraphics[width=3.2in,height=2.4in]{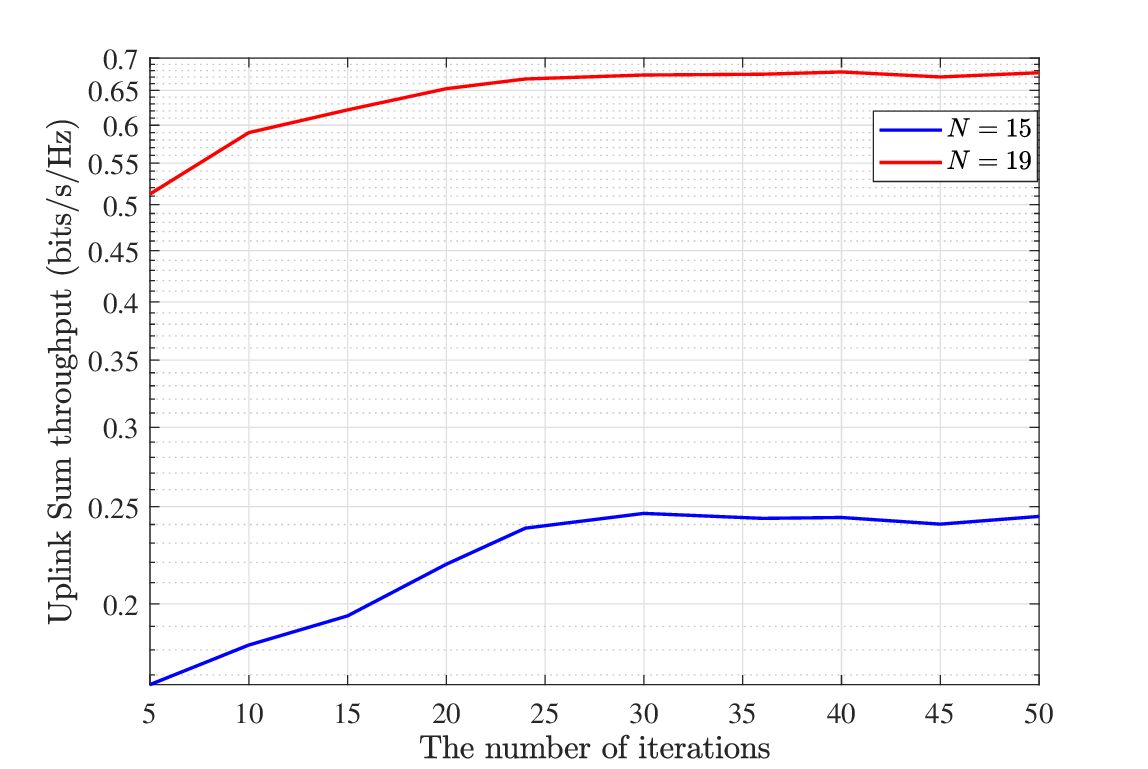}
\vspace{-0.3cm}
\caption{Optimized uplink sum throughput versus the number of iterations, where ${P_{{\rm{AP}}}}/{N_0} = 30$~dB, $\theta  = 0.1$, $R = 0.5$, and $JF{I^{th}} = 0.8$.}
\label{fig:convergence}
\vspace{-0.55cm}
\end{figure}

\begin{figure}
\center
\includegraphics[width=3.2in,height=2.4in]{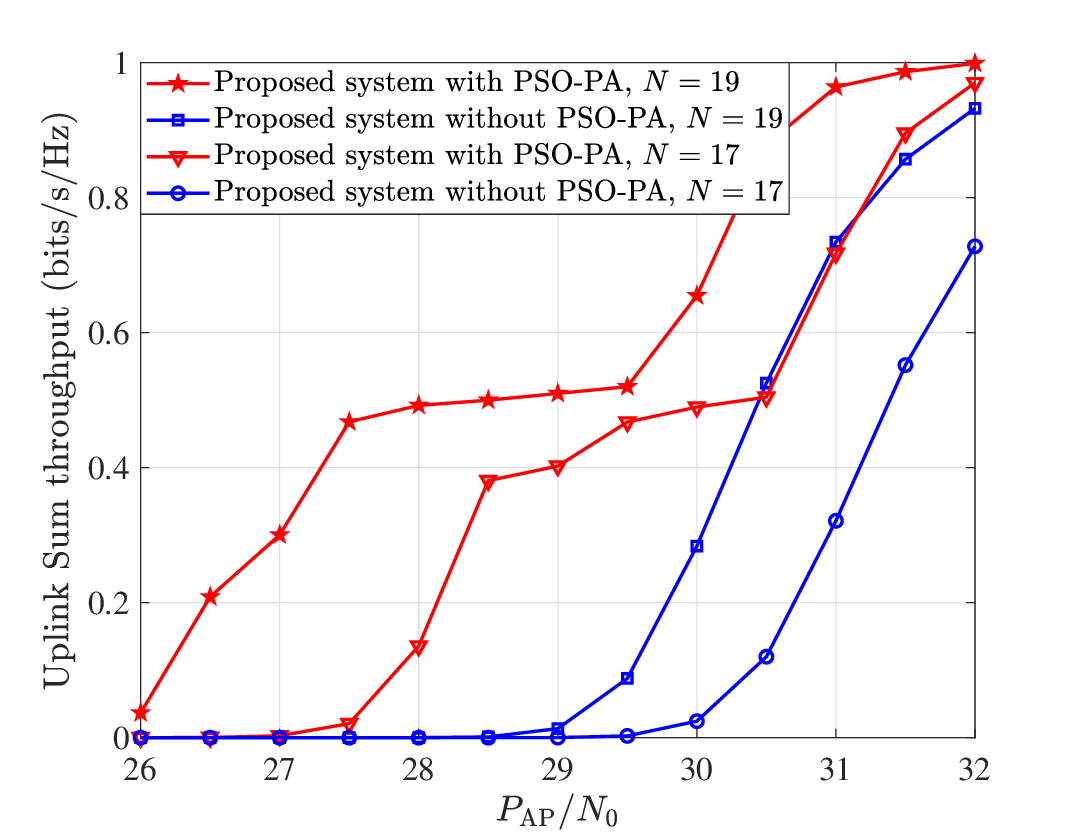}
\vspace{-0.3cm}
\caption{Optimized uplink sum throughput versus transmit SNR of the proposed system with various values of $N$, where $\theta  = 0.1$, $R = 0.5$, and $JF{I^{th}} = 0.8$.}
\label{fig:AoIVS}
\vspace{-0.55cm}
\end{figure}

Fig.~\ref{fig:convergence} shows the optimized uplink sum throughput versus the number of iterations, where ${P_{{\rm{AP}}}}/{N_0} = 30$~dB, $\theta  = 0.1$, $R = 0.5$, and $JF{I^{th}} = 0.8$. As can be seen in Fig.~\ref{fig:convergence}, the curves stabilize quickly with only a few iterations.
When $N$ goes large, convergence becomes faster, because it is easier to satisfy the constraints.
This indicates that PSO-PA is an effective method that can converge faster to obtain a solution.
Fig.~\ref{fig:AoIVS} plots the optimized uplink sum throughput versus transmit SNR of the proposed system with various values of $N$, where $\theta  = 0.1$, $R = 0.5$, and $JF{I^{th}} = 0.8$.
It is observed that the proposed PSO-PA method can improve the sum throughput performance of the proposed system.
For example, for a target sum throughput of $0.6$ bits/s/Hz, the proposed system with the PSO-PA method and $N = 19$ can obtain a $1$~dB gain compared with the proposed system without PSO-PA.
As the transmit SNR increases, the PSO-PA method can optimize the uplink sum throughput and meet the uplink joint outage probability and JFI requirements by adjusting the transmit power allocation factors and the STAR-RIS's energy-splitting ratios.

\vspace{-0.35cm}
\section{Conclusion} \label{sect:conclusion}
In this paper, we have investigated a STAR-RIS and energy buffer aided MISO SWIPT-NOMA system, in which the power transfer and information transmission states are modeled using Markov chains. Moreover, the power outage probability, information outage probability, and sum throughput expressions for the proposed system were derived in closed-form over Nakagami-$m$ fading channels.
From the perspective of power and information outage, the downlink and uplink joint outage probabilities were analyzed to investigate the overall network performance.
Both theoretical analysis and simulations showed that the proposed system yields an improved performance compared to the selected baseline schemes.
Furthermore, the PSO-PA algorithm was designed to maximize the uplink sum throughput with a constraint on the uplink joint outage probability and JFI.
Results showed that the proposed PSO-PA method significantly improves the uplink sum throughput performance of the proposed system.
Thanks to the aforementioned advantages, the proposed system could offer a promising candidate for large-scale, low-power, and sustainable IoT applications.

\vspace{-0.4cm}
\appendices
\section{}
Here, (\ref{eq:P11}) is derived.
The Bessel function of the first kind can be expanded as ${I_{\frac{q}{2} - 1}}\left( {\frac{s}{{{\sigma ^2}}}\sqrt x } \right) = \sum\limits_{\varsigma  = 0}^\infty  {\frac{{{{\left( {\frac{s}{{2{\sigma ^2}}}\sqrt x } \right)}^{\frac{q}{2} - 1 + 2\varsigma }}}}{{\varsigma !\Gamma \left( {\frac{q}{2} + \varsigma } \right)}}} .$

Thus, (\ref{eq:P11}) can be expressed as
\vspace{-0.20cm}
\begin{small}
\begin{align}
\label{eq:P11k}
&{P^k_{11}}\left( q \right) = \int_{\frac{{L\left( {q{E_s} - C} \right)}}{{\eta \left( {1 - \theta } \right){\beta _k }{P_{AP}}{T_s}}}}^\infty  {f_\mathcal{T}\left( x \right)} dx\nonumber\\
& = \int_{\frac{{L\left( {q{E_s} - C} \right)}}{{\eta \left( {1 - \theta } \right){\beta _k }{P_{AP}}{T_s}}}}^\infty  {\frac{1}{{2{\sigma ^2}}}{{\left( {\frac{x}{{{s^2}}}} \right)}^{\frac{{q - 2}}{4}}}{e^{ - \frac{{{s^2} + x}}{{2{\sigma ^2}}}}}{I_{\frac{q}{2} - 1}}\left( {\frac{s}{{{\sigma ^2}}}\sqrt x } \right)} dx\nonumber\\
& \!=\! \sum\limits_{\varsigma = 0}^\infty  {\frac{{{e^{ - \frac{{{s^2}}}{{2{\sigma ^2}}}}}{s^{2\varsigma}}}}{{{{\left( {2{\sigma ^2}} \right)}^{\frac{q}{2} + 2\varsigma}}\varsigma!\Gamma \left( {\frac{q}{2} + \varsigma} \right)}}\!\!\int_{\frac{{L\left( {q{E_s} - C} \right)}}{{\eta \left( {1 - \theta } \right){\beta _k }{P_{AP}}{T_s}}}}^\infty \!\!\!\!\!\! {{e^{ - \frac{x}{{2{\sigma ^2}}}}}{x^{\frac{q}{2} - 1 + \varsigma}}} } dx.
\vspace{-0.20cm}
\end{align}
\end{small}%

From (3.351.2) in \cite{Gradshte1965Table}, one has
\vspace{-0.20cm}
\begin{small}
\begin{align}
\label{eq:3.351}
\int\limits_{\bar u}^\infty  {{x^{\omega '}}{e^{ - \bar \mu x}}dx = {e^{ - \bar u\bar \mu }}} \sum\limits_{\bar \ell  = 0}^{\omega '} {\frac{{\omega '!}}{{\bar \ell !}}} \frac{{{{\bar u}^{\bar \ell }}}}{{{{\bar \mu }^{\omega ' - \bar \ell  + 1}}}}.
\vspace{-0.4cm}
\end{align}
\end{small}%
Using (\ref{eq:3.351}), (\ref{eq:P11k}) can be calculated as
\vspace{-0.20cm}
\begin{small}
\begin{align}
&{P^k_{11}}\left( q \right) = \sum\limits_{\varsigma = 0}^\infty  {\frac{{{e^{ - \frac{{{s^2}}}{{2{\sigma ^2}}}}}{s^{2\varsigma}}}}{{{{\left( {2{\sigma ^2}} \right)}^{\frac{q}{2} + 2\varsigma}}\varsigma!\Gamma \left( {\frac{q}{2} + \varsigma} \right)}}} {e^{ - \frac{{L\left( {q{E_s} - C} \right)}}{{2{\sigma ^2}\eta \left( {1 - \theta } \right){\beta _k }{P_{\rm{AP}}}{T_s}}}}}\nonumber\\
 &\times \sum\limits_{p = 0}^{^{\frac{q}{2} - 1 + \varsigma}} {\frac{{\left( {\frac{q}{2} - 1 + \varsigma} \right)!{{\left( {\frac{{L\left( {q{E_s} - C} \right)}}{{\eta \left( {1 - \theta } \right){\beta _k }{P_{\rm{AP}}}{T_s}}}} \right)}^p}}}{{p!{{\left( {\frac{1}{{2{\sigma ^2}}}} \right)}^{\frac{q}{2} + \varsigma - p}}}}} \nonumber\\
&\!=\!\! \sum\limits_{\varsigma  = 0}^\infty  \!\!{\sum\limits_{p = 0}^{^{\frac{q}{2} - 1 + \varsigma}}\!\!\!\! {\frac{{{s^{2\varsigma }}{e^{ - \frac{{{s^2}}}{{2{\sigma ^2}}}}}}}{{{{\left( {2{\sigma ^2}} \right)}^{\varsigma  + p}}\varsigma !}}} } {e^{ - \frac{{L\left( {q{E_s} - C} \right)}}{{2{\sigma ^2}\eta \left( {1 - \theta } \right){\beta _k }{P_{\rm{AP}}}{T_s}}}}}\frac{{{{\!\left( {\frac{{L\left( {q{E_s} - C} \right)}}{{\eta \left( {1 - \theta } \right){\beta _k }{P_{\rm{AP}}}{T_s}}}} \!\right)}^p}}}{{p!}}.
\vspace{-0.20cm}
\end{align}
\end{small}%
\vspace{-0.85cm}
\section{}
Here, (\ref{eq:P10u}) is derived.
\vspace{-0.20cm}
\begin{small}
\begin{align}
&P_{10,u}^{t,io} = \Pr \left( {\gamma _s^u < {\gamma _{th}}} \right)\nonumber\\
& = \Pr \left( \begin{array}{l}
{\left| {\sum\limits_{\ell  = 1}^L {\left( {h_{t,\ell }^0 + \sum\limits_{i = 1}^N {{h_{\ell ,i}}{g_{t,i}}} } \right)} } \right|^2} < \\
\frac{{{P_r}{\beta _r}{{\left| {\sum\limits_{\ell  = 1}^L {\left( {h_{r,\ell }^0 + \sum\limits_{i = 1}^N {{h_{\ell ,i}}{g_{r,i}}} } \right)} } \right|}^2}{\gamma _{th}} + L{N_0}{\gamma _{th}}}}{{{P_t}{\beta _t}}}
\end{array} \right)\nonumber\\
& = \Pr \left( {{\cal Z}_{l,t}^2 < \frac{{{P_r}{\beta _r}{\cal Z}_{l,r}^2{\gamma _{th}} + L{N_0}{\gamma _{th}}}}{{{P_t}{\beta _t}}}} \right).
\vspace{-0.20cm}
\end{align}
\end{small}%
From (\ref{eq:CDFtP1}) and (\ref{eq:PDFtP1}), one has
\vspace{-0.20cm}
\begin{small}
\begin{align}
\label{eq:P10utio}
&P_{10,u}^{t,io} = {F_{{\cal Z}_{l,t}^2}}\left( {\frac{{{P_r}{\beta _r}{\cal Z}_{l,r}^2{\gamma _{th}} + L{N_0}{\gamma _{th}}}}{{{P_t}{\beta _t}}}} \right)\nonumber\\
& = \int\limits_0^\infty  {{F_{{\cal Z}_{l,t}^2}}\left( {\frac{{{P_r}{\beta _r}x{\gamma _{th}} + L{N_0}{\gamma _{th}}}}{{{P_t}{\beta _t}}}} \right)} {f_{{\cal Z}_{l,r}^2}}\left( x \right)dx\nonumber\\
& = 1 - \sum\limits_{m = 0}^{L\chi  - 1}  \frac{{{\phi ^{L\chi  + m}}}}{{2m!\left( {L\chi  - 1} \right)!}}\int\limits_0^\infty   {e^{ - \phi {{\left( {\frac{{{P_r}{\beta _r}x{\gamma _{th}} + L{N_0}{\gamma _{th}}}}{{{P_t}{\beta _t}}}} \right)}^{\frac{1}{2}}}}}\nonumber\\
&\times {\left( {\frac{{{P_r}{\beta _r}x{\gamma _{th}} + L{N_0}{\gamma _{th}}}}{{{P_t}{\beta _t}}}} \right)^{\frac{m}{2}}}{e^{ - \phi {x^{\frac{1}{2}}}}}{x^{\frac{{L\chi  - 2}}{2}}}dx.
\vspace{-0.20cm}
\end{align}
\end{small}%

Equation (\ref{eq:P10utio}) can be evaluated by leveraging the Gauss-Hermite quadrature approach. According to Table (25.10) in \cite{1966Handbook}, one has $\int_{ - \infty }^\infty  {g\left( x \right)} dx = \sum\limits_{w = 1}^W {{\psi _w}} g\left( {{x_w}} \right)\exp \left( {x_w^2} \right) + {O_W}$,
where $W$ denotes the number of sample points used for approximation, ${{x_w}}$ is the $w$-th root of the Hermite polynomial ${H_W}(x)$~$(w = 1,2, \ldots W)$, ${{\psi  _w}}$ is the $w$-th associated
weight obtained from $\frac{{{2^{W - 1}}W!\sqrt \pi  }}{{{W^2}H_{W - 1}^2({x_w})}}$ and ${O_W}$ is the residual term that tends to $0$ when $W$ tends to infinity.

Using variable substitution $\ln ({x}) = u$ to obtain the new limits with integral from $ - \infty $ to $ + \infty $ and after some mathematical operations, one obtains
\vspace{-0.20cm}
\begin{small}
\begin{align}
\label{eq:ACP10}
&P_{10,u}^{t,io}\!\approx\! 1 \!-\!\! \sum\limits_{m = 0}^{L\chi  - 1} \!\! \frac{{{\phi ^{L\chi  + m}}}}{{2m!\left( {L\chi  - 1} \right)!}}\!\sum\limits_{w = 1}^W \!{{\psi _w}} {e^{ - \phi {{\left( {\frac{{{P_r}{\beta _r}{e^{{u_w}}}{\gamma _{th}} + L{N_0}{\gamma _{th}}}}{{{P_t}{\beta _t}}}} \!\right)}^{\!\!\frac{1}{2}}}}}\nonumber\\
& \!\times \!\!{\left(\!\! {\frac{{{P_r}{\beta _r}{e^{{u_w}}}{\gamma _{th}} \!+\! L{N_0}{\gamma _{th}}}}{{{P_t}{\beta _t}}}} \!\!\right)^{\!\!\frac{m}{2}}}\!\!\!\!{e^{ - \phi {e^{\frac{{{u_w}}}{2}}}}}\!\!{e^{{u_w}\frac{{L\chi  - 2}}{2}}}\!{e^{{u_w} + u_w^2}}\! - \!{O_W}.\!\!\!
\vspace{-0.20cm}
\end{align}
\end{small}%

Moreover, (\ref{eq:P10ur}) is derived as:
\vspace{-0.20cm}
\begin{small}
\begin{align}
\label{eq:ACP10r}
&P_{10,u}^{r,io} = 1 - \Pr \left( {\gamma _s^u > {\gamma _{th}},\gamma _r^u > {\gamma _{th}}} \right)\nonumber\\
& = 1 - \Pr \left( {\frac{{{\gamma _{th}}L{N_0}}}{{{P_r}{\beta _r}}} < {\cal Z}_{l,r}^2 < \frac{{{P_t}{\beta _t}{\cal Z}_{l,t}^2 - {\gamma _{th}}L{N_0}}}{{{\gamma _{th}}{P_r}{\beta _r}}}} \right)\nonumber\\
& = 1 - {F_{{\cal Z}_{l,r}^2}}\left( {\frac{{{P_t}{\beta _t}{\cal Z}_{l,t}^2 - {\gamma _{th}}L{N_0}}}{{{\gamma _{th}}{P_r}{\beta _r}}}} \right) + {F_{{\cal Z}_{l,r}^2}}\left( B\right),
\vspace{-0.3cm}
\end{align}
\end{small}%
where $B = \frac{{{\gamma _{th}}L{N_0}}}{{{P_r}{\beta _r}}}$.

Similar to (\ref{eq:ACP10}), one has
\vspace{-0.15cm}
\begin{small}
\begin{align}
\label{eq:ACP10r1}
&{F_{{\cal Z}_{l,r}^2}}\left( {\frac{{{P_t}{\beta _t}{\cal Z}_{l,t}^2 - {\gamma _{th}}L{N_0}}}{{{\gamma _{th}}{P_r}{\beta _r}}}} \right)\nonumber\\
& = \int\limits_A^\infty  {{F_{{\cal Z}_{l,r}^2}}\left( {\frac{{{P_t}{\beta _t}x - {\gamma _{th}}L{N_0}}}{{{\gamma _{th}}{P_r}{\beta _r}}}} \right)} {f_{{\cal Z}_{l,t}^2}}\left( x \right)dx\nonumber\\
& = 1 \!-\! {F_{{\cal Z}_{l,t}^2}}\left( A \right) \!-\! \sum\limits_{m = 0}^{L\chi  - 1}  \frac{{{\phi ^{L\chi  + m}}}}{{2m!\left( {L\chi  - 1} \right)!}}\int\limits_A^\infty  {{e^{ - \phi {{\left( {\frac{{{P_t}{\beta _t}x - {\gamma _{th}}L{N_0}}}{{{\gamma _{th}}{P_r}{\beta _r}}}} \right)}^{\frac{1}{2}}}}}} \nonumber\\
& \times {\left( {\frac{{{P_t}{\beta _t}x - {\gamma _{th}}L{N_0}}}{{{\gamma _{th}}{P_r}{\beta _r}}}} \right)^{\frac{m}{2}}}{e^{ - \phi {x^{\frac{1}{2}}}}}{x^{\frac{{L\chi  - 2}}{2}}}dx\nonumber\\
& \approx 1 - {F_{{\cal Z}_{l,t}^2}}\left( A \right) - \sum\limits_{m = 0}^{L\chi  - 1}  \frac{{{\phi ^{L\chi  + m}}}}{{2m!\left( {L\chi  - 1} \right)!}}\sum\limits_{w = 1}^W {{\psi _w}} \nonumber\\
& \!\times\! {e^{ - \phi {{\left( {\frac{{{P_t}{\beta _t}\left( {{e^{{u_w}}} + A} \right) - {\gamma _{th}}L{N_0}}}{{{\gamma _{th}}{P_r}{\beta _r}}}} \right)}^{\frac{1}{2}}}}}\!\!{\left( {\frac{{{P_t}{\beta _t}\left( {{e^{{u_w}}} + A} \right) - {\gamma _{th}}L{N_0}}}{{{\gamma _{th}}{P_r}{\beta _r}}}} \right)^{\frac{m}{2}}}\nonumber\\
& \times {e^{ - \phi {{\left( {{e^{{u_w}}} + A} \right)}^{\frac{1}{2}}}}}{\left( {{e^{{u_w}}} + A} \right)^{\frac{{L\chi  - 2}}{2}}}{e^{{u_w} + u_w^2}} - {O_W},
\vspace{-0.30cm}
\end{align}
\end{small}%
where $A = \frac{{{\gamma _{th}}L{N_0}}}{{{P_t}{\beta _t}}}$.
By combining (\ref{eq:ACP10r}) and (\ref{eq:ACP10r1}), (\ref{eq:P10ur}) is obtained.

\vspace{-0.7cm}


\end{document}